\documentclass[twocolumn,aps]{revtex4}
\usepackage{graphics}
\usepackage{bm}
\usepackage[fleqn]{amsmath}
\usepackage{hyperref}
\usepackage{exscale,relsize}
\usepackage{epsfig}
\usepackage{amssymb}
\usepackage{amsmath}
\usepackage{euscript}
\usepackage{float}
 \usepackage{amsthm}

\newcommand{\be}{\begin{equation}}
\newcommand{\e}{\end{equation}}

\newcommand{\beml}{\begin{subequations}}
\newcommand{\eml}{\end{subequations}}
\newcommand{\beq}{\begin{eqnarray}}
\newcommand{\eq}{\end{eqnarray}}
\newcommand{\ba}{\begin{array}}
\newcommand{\ea}{\end{array}}

\newcommand{\lt}{\left}
\newcommand{\rt}{\right}
\newcommand{\n}{\nonumber}
\newcommand{\vs}{\vec{\sigma}}
\newcommand{\s}{\sigma}
\newcommand{\la}{\langle}
\newcommand{\ra}{\rangle}
\newcommand{\tr}{{\rm Tr}\,}

\newcommand{\re}{\,{\rm Re}\,}

\newcommand{\Dp}{\Delta^{(+)}}
\newcommand{\Dm}{\Delta^{(-)}}

\newcommand{\ur}{\urcorner}
\newcommand{\llc}{\llcorner}

\begin{document}

\date{\today}
\title{Rigorous derivation of the triple scattering signal from single-atom responses}

\author{Vyacheslav Shatokhin and Thomas Wellens}
\affiliation{Institute of Physics, University of Freiburg, Hermann-Herder-Str. 3,
D-79104 Freiburg, Germany}

\begin{abstract}
We study coherent backscattering of intense laser light from
three immobile two-level atoms using the master equation approach.
The master equation is solved analytically, and the triple
scattering spectrum is expressed in quadratures. We show that this solution can be obtained via a self-consistent combination of single-atom spectral
responses, and is equivalent to the solution following from the diagrammatic pump-probe approach. We deduce the general expressions for
single-atom building blocks which can be used
in simulations of multiple inelastic scattering of laser light in the bulk atomic medium.

\end{abstract}
%

\maketitle

\section{Introduction}
Coherent backscattering (CBS) is an enhancement of the average
intensity of the resonant wave backscattered off a dilute disordered
medium due to the constructive interference between the
counter-propagating multiply scattered amplitudes possessing
time-reversal symmetry \cite{albada85}. Recent experiments on CBS of
light from cold atoms
\cite{labeyrie99,kulatunga03,labeyrie03,chaneliere04,balik05}
studied the impact of the atomic internal structure on the phase
coherence of the interfering waves. It was shown that the spin-flips
of a photon on degenerate dipole transitions of rubidium atoms
\cite{labeyrie99,kulatunga03,labeyrie03,balik05}, as well as
nonlinear inelastic scattering from saturated strontium
\cite{chaneliere04} and rubidium \cite{balik05} atoms induced by
intense laser field significantly reduce the enhanced
backscattering. It should be mentioned that, besides the importance of its own, the studies of CBS of light by cold atoms are also motivated by the interest in photon localization \cite{akkermans08,sokolov09} and random lasing \cite{savels07,goetschy11} in dense atomic clouds.

Whereas there is a detailed description of CBS of light by atoms with
arbitrary internal degeneracy in the elastic (linear) scattering regime
\cite{labeyrie03,mueller02,kupriyanov03}, multiple inelastic scattering theory
exists only for atoms weakly saturated by two photons
\cite{wellens06}. But the parameter regimes of moderate or strong saturations
probed in the experiments \cite{chaneliere04,balik05} requires a non-perturbative
treatment of the atom-laser field interactions, and is beyond the reach of the
diagrammatic approach put forward in \cite{wellens06}.

Standard non-perturbative quantum-optical methods, such as a master
equation approach (see, for instance,
\cite{lehmberg70,agarwal_book,cohen-tannoudji}), can be used for a
description of CBS of intense laser light from cold atoms. However,
due to the exponential growth of the Hilbert space with increasing
number of scatterers these methods are fundamentally limited to
systems of a few scatterers. So far, the master equation approach
and extensions thereof have been successfully applied to treat
double scattering
\cite{shatokhin05,shatokhin06,gremaud06,shatokhin07}.

To avoid the above problem of the exponential growth of the Hilbert
space, and yet to account for the atom-laser field interaction
strength non-perturbatively, we proposed a stochastic method of
solving the multiple scattering problem in dilute cold atomic clouds
which we call the pump-probe approach to CBS
\cite{wellens10,geiger10}. The main idea of this method is to
express the multiple scattering signal through single-atom responses
to a classical polychromatic random field, of which one component
represents a laser (pump) field, while the remaining components describe
weak (probe) fields scattered from the surrounding distant atoms.

So far, the validity of the pump-probe approach was
proven in the case of double
scattering by two two-level atoms \cite{wellens10,geiger10}  by analytically establishing the equivalence with the results
following from the master equation approach \cite{shatokhin10}.
Furthermore, we showed that the expressions for the double
scattering background and interference spectra can be represented
graphically, by self-consistently combining single-atom building
blocks \cite{shatokhin12}. Thus, the pump-probe approach is related
to the non-perturbative methods of quantum optics, on the one hand,
and to the diagrammatic scattering theories, on the other hand.

In the present contribution, we explore these relations further.
Namely, from the rigorous solutions of the master equation for three
two-level atoms we deduce the analytical expressions for the triple
scattering background and interference spectra of CBS. We present
our results diagrammatically by
 self-consistently combining single-atom building blocks, analogously to the case of double scattering \cite{shatokhin12}.  Thereby, we show the equivalence between
 the master equation and the pump-probe approaches to CBS for triple scattering.
Moreover, we deduce general expressions describing the spectral response of an atom subjected to an arbitrary number of probe fields. These expressions will be required in future work to implement the pump-probe approach for a medium consisting of an arbitrarily large number of atoms.

The structure of this paper is as follows. In the next section, we present the model,
the master equation, and the main quantity of interest -- the triple scattering
contribution to CBS of light from three two-level atoms. In Sec.~\ref{sec:solving_meq}
we describe the method to find an analytical solution of the master equation which links it to the pump-probe approach.
Sections~\ref{sec:results} and \ref{sec:expgen} contain our main results. They include
diagrammatic expansions of single-atom building blocks into the elastic and inelastic spectral responses whose composition yields the ladder and crossed
spectra (Sec.~\ref{sec:results}), and the general expressions for single-atom building
blocks (Sec.~\ref{sec:expgen}). We present analytical and numerical triple scattering
spectra in Sec.~\ref{sec:example}. Conclusions of this work are presented in Sec.~\ref{sec:discuss}.

\section{Model}
\label{sec:formalism}
\subsection{Master equation}
\begin{figure}
\includegraphics[width=5cm]{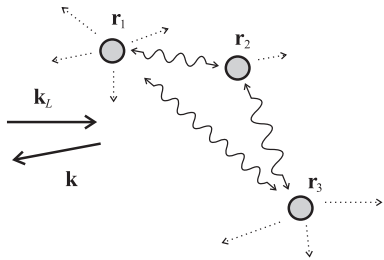}
\caption{Three atoms (gray circles) located in the radiation zone of each other are driven by a quasi-resonant laser field of
arbitrary intensity with a wave-vector ${\bf k}_L$. They radiatively decay (dotted arrows)  into a common electromagnetic bath, which also
mediates the resonant dipole-dipole interaction (double wavy
arrows). We will be interested in the
far-field intensity of the scattered light along ${\bf k}$.}
\label{fig:model}
\end{figure}

We will start out with introducing our model
depicted in Fig.~\ref{fig:model}.
It includes three quantum scatterers (two-level atoms) embedded in a
common electromagnetic reservoir at fixed random positions ${\bf r}_\lambda$ and
excited by a quasi-resonant continuous wave laser field with a wave vector ${\bf
k}_L$. We will be interested in the disorder-averaged far-field intensity of the light scattered into a given direction ${\bf k}$.

To describe CBS of light from three laser-driven atoms, we will employ
the master equation approach \cite{lehmberg70,agarwal_book} which
has already been used for
calculating double scattering from two saturated Sr
\cite{shatokhin05,shatokhin07} and two-level atoms
\cite{shatokhin10}.
Therefore, we will next present a
master equation
governing the evolution of the quantum mechanical
expectation values of an arbitrary atomic observable $Q$ of a
three-atom system. In the Heisenberg picture and in the frame
rotating at the laser frequency $\omega_L$, $\la Q\ra$ obeys the
following equation of motion \cite{shatokhin10}:
 \begin{align}
\langle\dot{Q}\rangle&=\sum_{\lambda=1}^3\langle-i\delta[\sigma^+_\lambda\sigma^-_\lambda,Q]
-\frac{i}{2}[\Omega_\lambda\sigma_\lambda^++\Omega^*_\lambda \sigma^-_\lambda,Q]\n\\
&-\gamma(\sigma^+_\lambda\sigma^-_\lambda Q+Q\sigma^+_\lambda\sigma^-_\lambda-2\sigma^
+_\lambda Q\sigma^-_\lambda)\rangle
\label{meq}\\
&+\sum_{\lambda\neq \mu=1}^3\left(T_{\lambda\mu}\langle[\sigma_\lambda^+Q,\sigma^-_\mu]
+~T_{\lambda\mu}^*[\sigma^+_\lambda, Q\sigma^-_\mu]\rangle\right).\n
\end{align}
Here, $\sigma^-_\lambda=|1\rangle_\lambda\langle 2|_\lambda$ and
$\sigma_\lambda^+=|2\rangle_\lambda\langle 1|_\lambda$, with $|1\rangle_\lambda$ and
$|2\rangle_\lambda$ being respectively the ground and excited states of
atom $\lambda$, denote the atomic lowering and raising operators.
Furthermore, $\Omega_{\lambda}=\Omega e^{i{\bf k}_L\cdot{\bf r}_\lambda}$ is the
Rabi frequency dependent on the atomic position ${\bf r}_\lambda$,
$\delta=\omega_L-\omega_0$ is the laser-atom detuning, and $\gamma$
is half the radiative decay rate of the excited state.  The lower
line of Eq.~(\ref{meq}) describes the retarded dipole-dipole
interaction proportional to the complex couplings $T_{\lambda\mu}$. The couplings $T_{\lambda\mu}$, in turn, are
inversely proportional to the distance between the atoms $\lambda$ and
$\mu$: \be T_{\lambda\mu}\equiv
i\frac{3}{2}\gamma(1-(\hat{\bf d}\cdot\hat{\bf r}_{\lambda\mu})^2)\frac{e^{-ik_Lr_{\lambda\mu}}}{k_Lr_{\lambda\mu}}, \label{dipdip}
\e where $\hat{\bf d}$ and $\hat{\bf r}_{\lambda\mu}$ are unit vectors along the atomic dipole moment and radius-vector connecting the atoms $\lambda$ and $\mu$, respectively, and $r_{\lambda\mu}= |{\bf r}_\lambda-{\bf r}_\mu|$. It should be stressed that
Eq.~(\ref{dipdip}) is valid only in
 the limit $k_Lr_{\lambda\mu}\gg 1$, which is standard in the
 context of CBS, and is also assumed here. Note that although the constants
 $T_{\lambda\mu}=T_{\mu\lambda}$ and
$T_{\lambda\mu}^*=T_{\mu\lambda}^*$, the interaction Liouvillians in Eq.~(\ref{meq}) associated with these constants are not symmetric under the permutation of the indices
$\lambda$ and $\mu$. Physically, they describe different excitation transfer
processes induced by the far-field
 dipole-dipole interaction Liouvillian \cite{shatokhin10}. We will discuss this issue in Sec.~\ref{sec:inter_matr}.

\subsection{Triple scattering intensity}
Solution of Eq.~(\ref{meq}) gives access to the expectation values of
observables of the field radiated by a system of three atoms. We will
focus on the triple scattering contribution to the
stationary scattered light intensity in the direction
of the wave vector ${\bf k}$. This quantity is a part of the general
expression for the intensity of the scattered light given by \be \langle I({\bf
k})\rangle_{\rm ss}=\sum_\lambda\la\s^+_\lambda\s^-_\lambda\ra_{\rm ss}+\sum_{\lambda,\mu\neq
\lambda}\la\s_\lambda^+\otimes\s^-_\mu\ra_{\rm ss} e^{i{\bf k}\cdot({\bf r}_\lambda-{\bf
r}_\mu)} \label{eq:in2}, \e where the angular brackets
$\la\ldots\ra_{\rm ss}$ denote the quantum-mechanical average in the
steady state. The existence of the steady state is guaranteed by the fact that the corresponding expectation values in the right hand side of Eq.~(\ref{eq:in2}) are deduced from the solution of a linear equation (see Eq.~(\ref{mateq}) below) with a non-singular evolution matrix $(A+V)$ whose eigenvalues have negative real parts.
The first and second terms in Eq.~(\ref{eq:in2}) describe the incoherent (or background) and the interference intensities,
respectively.

In the framework of the master equation approach, the multiple scattering contributions correspond to the subsequent terms in a perturbative expansion of the stationary solution of Eq.~(\ref{meq}) in the power series of the dipole-dipole interaction constants. The first-order term, proportional to $|T_{\lambda\mu}|$, describes {\it the amplitude of double scattering} between the atoms $\lambda$ and $\mu$. Consequently, {\it the intensity of triple scattering}  between atoms $1$, $2$ and $3$ must be proportional to $|T_{12}T_{23}|^2$. Furthermore, the CBS signal is observed after the configuration, or disorder, averaging
over the random positions of the scatterers, which will be denoted as
$\la\ldots\ra_{\rm conf}$. Summarizing the above, we can write
the expression for the triple scattering contribution to CBS as follows:
\begin{align}\la\langle I({\bf k})\rangle_{\rm ss}^{(4)}\ra_{\rm conf}&=\sum_{\lambda=1}^{3}\la\la\s^+_\lambda\s^-_\lambda\ra_{\rm ss}^{(4)}\ra_{\rm conf}\label{eq:CBS_Int}\\
&+\sum_{\lambda,\mu\neq \lambda}^3\la\la\s_\lambda^+\otimes\s^-_\mu\ra_{\rm ss}^{(4)} e^{i{\bf k}\cdot({\bf r}_\lambda-{\bf r}_\mu)}\ra_{\rm conf},\n
\end{align}
where the superscript $(4)$ denotes the fourth order in the dipole-dipole coupling. Following the standard nomenclature in the field of coherent quantum transport \cite{sheng}, we will refer to the configuration averaged background and interference contributions as the ladder and crossed intensities, respectively.

In the following, we will ignore recurrent scattering, which is proportional to $|T_{\lambda\mu}|^4$. Although in a three-atom master equation, the recurrent and triple scattering contributions are of the same order of magnitude, in a dilute atomic cloud, the impact of the recurrent scattering on the observed signal scales as $N^2$, and is by far dominated, for $N\gg 1$, by that of the triple scattering which scales as $N^3$.

We will identify all triple scattering contributions which survive the disorder averaging. Since some of them are equal to each other, it will be sufficient to find only the non-equivalent contributions.
To this end, in the next section we will study the structure and solutions of the master equation (\ref{meq}) in more detail.

\section{Solution of the master equation}
\label{sec:solving_meq}
\subsection{Basis set of operators}
Let us consider the evolution
of the expectation value of an operator $Q$ from the complete
three-atom basis set of operators: $Q\in \{\vec{q}_1\otimes \vec{ q}_2\otimes\vec{ q}_3\}$,
with \beq \vec{ q}_\lambda&=&(\openone_\lambda,\sigma^-_\lambda,\sigma^+_\lambda,\sigma^z_\lambda)^T,\n\\
\openone_\lambda&=&\sigma^+_\lambda\sigma^-_\lambda+\sigma^-_\lambda\sigma^+_\lambda, \quad
\sigma^z_\lambda=\sigma^+_\lambda\sigma^-_\lambda-\sigma^-_\lambda\sigma^+_\lambda.\eq
The normalization condition $\tr[ \openone_1\otimes
\openone_2\otimes\openone_3\rho]=1$ which holds for any atomic density operator
$\rho$ reduces the size of the operator space by one, leading to a set of 63 operators. Their expectation values will be ordered
as elements of the vector \be
\la\vec{ Q}\ra=(\vec{x},\vec{y},\vec{z})^T,
\label{vecQ}\e with
\beml
\begin{align}
\vec{x}&=(\la\vec{\s}_3\ra,\la\vec{\s}_2\ra,\la\vec{\s}_1\ra)^T,\\
\vec{y}&=(\la\vec{\s}_2\otimes\vec{\s}_3\ra,\la\vec{\s}_1\otimes\vec{\s}_3\ra,\la\vec{\s}_1\otimes\vec{\s}_2\ra)^T,\\
\vec{z}&=\la\vec{\s}_1\otimes\vec{\s}_2\otimes\vec{\s}_3\ra,
\end{align}
\label{eq:xyz} \eml where $\la\vec{\s}_\lambda\ra$ is a Bloch vector corresponding to atom
$\lambda$: \be \la\vec{\s}_\lambda\ra=(\la\s^-_\lambda\ra,\la\s^+_\lambda\ra,\la\s^z_\lambda\ra)^T.\label{eq:Bloch}\e

The evolution of the vector $\la\vec{Q}\ra$ is governed by the equation of motion: \be
\la\dot{\vec{ Q}}\ra=(A+V)\la\vec{Q}\ra+\vec{\Lambda},\label{mateq}\e
where the matrices $A$ and $V$ govern the dynamics of
independent and dipole-dipole interacting atoms, respectively, and
the vector $\vec{ \Lambda}$ is given by Eq.~(\ref{eq:nullvector}) in Appendix \ref{sec:appA}.

The steady state solution of Eq.~(\ref{mateq}) which is of the
fourth order in the matrix $V$ reads:
\be
\la\vec{Q}\ra_{\rm ss}^{(4)}=(GV)^4G\vec{\Lambda},
\label{eq:Q4}
\e
where $G\equiv -A^{-1}$.  Given the matrices $A$, $V$, and $\vec{\Lambda}$, it is easy to find $\la\vec{Q}\ra_{\rm ss}^{(4)}$ numerically for any random positions ${\bf r}_i$ of the atoms. Averaging this result over disorder gives $\la\la\vec{Q}^{(4)}\ra_{\rm ss}\ra_{\rm conf}$ and, consequently, the terms from the right hand side of Eq.~(\ref{eq:CBS_Int}) contributing to the ladder and crossed intensities. However,
in this work we would like to prove the equivalence
of the results of the master equation and the pump-probe approaches. Therefore, we will adhere to the analytical tools developed in \cite{shatokhin10,shatokhin12}. As will be seen below in Sec.~\ref{sec:expgen}, using the analytical methods will enable us not only to prove the equivalence between the two methods, but also to deduce the general expressions for the single-atom building blocks which will be required in the future treatment of the multiple scattering of laser light in cold atomic clouds.

We will proceed with discussing the structure of the evolution matrices and presenting the recurrence relations connecting the steady-state vectors $\vec{x}$, $\vec{y}$, and $\vec{z}$.

\subsection{Recurrence relations}
\label{sec:recurr}

For a vector $\la\vec{Q}\ra$ given by Eq.~(\ref{vecQ}), the matrices $A$ and $V$ have the following block structure (see Appendix \ref{sec:appA}): \be
A=\lt(\ba{ccc}M_+&0&0\\
L_+&M_\times&0\\
0&L_\times&N_\times\ea\rt),\quad V=\lt(\ba{ccc}0&U_\ur&0\\
U_\llc&U_\times&W_\urcorner\\
0&W_\llcorner& W_\times\ea\rt).
\label{eq:matr_A_V}
\e
A perturbative expansion of the steady-state solution of Eq.~(\ref{mateq}) in the power series of the matrix $V$ yields a system of coupled recurrence relations for the vectors $\vec{x}^{\;(n)}$, $\vec{y}^{\;(n)}$, and $\vec{z}^{\;(n)}$:
\beml
\begin{align}
\vec{x}^{\;(n)}&=G_+U_\ur\vec{y}^{\;(n-1)},\\
\vec{y}^{\;(n)}&=G_\times U_\llc\vec{x}^{\;(n-1)}+G_\times U_\times\vec{y}^{\;(n-1)}
+G_\times L_+\vec{x}^{\;(n)}\n\\&+G_\times W_\ur\vec{z}^{\;(n-1)},\label{eq:y^n}\\
\vec{z}^{\;(n)}&=D_\times W_\llc\vec{y}^{\;(n-1)}+
D_\times W_\times\vec{z}^{\;(n-1)}+
D_\times L_\times\vec{y}^{\;(n)},\label{eq:z^n}
\end{align}
\label{eq:recurr} \eml with the initial conditions determined by
$\la\vec{Q}\ra^{(n)}_{\rm ss}=0$ for $n<0$, and
$\la\vec{Q}\ra^{(0)}_{\rm
ss}=(\vec{x}^{\;(0)},\vec{y}^{\;(0)},\vec{z}^{\;(0)})^T$, where \beml
\begin{align}
\vec{x}^{\;(0)}=(&\la\vs_3\ra^{(0)},\la\vs_2\ra^{(0)},\la\vs_1\ra^{(0)})^T,\\
\vec{y}^{\;(0)}=(&\la\vs_2\ra^{(0)}\otimes \la\vs_3\ra^{(0)},
\la\vs_1\ra^{(0)}\otimes \la\vs_3\ra^{(0)},\n\\
&\la\vs_1\ra^{(0)}\otimes \la\vs_2\ra^{(0)})^T,\label{eq:init_y^0}\\
\vec{z}^{\;(0)}=\la&\vs_1\ra^{(0)}\otimes \la\vs_2\ra^{(0)}\otimes \la\vs_3\ra^{(0)},\label{eq:init_z0}
\end{align}
\label{eq:init_cond}
\eml
and $\la\vs_\lambda\ra^{(0)}=G_\lambda\vec{L}$.
In
Eqs.~(\ref{eq:recurr}) and (\ref{eq:init_cond}), we introduced a shorthand notation $G_\lambda\equiv G_\lambda(0)$ ($\lambda=1,2,3$),
$G_\alpha\equiv G_\alpha(0)$ ($\alpha=+,\times$), $D_\times\equiv D_\times(0)$, where
\be
G_\lambda(z)=\frac{1}{z-M_\lambda},\;
G_\alpha(z)=\frac{1}{z-M_\alpha},\;
D_\times(z)=\frac{1}{z-N_\times}\label{eq:Greens},\e are the Green's matrices.

The possibility of demonstrating the equivalence between the results of the master equation and the pump-probe approach is based on the analytical solution of the recurrence relations (\ref{eq:recurr}) for fixed atomic positions, and on the subsequent analytical configuration averaging procedure.

In the following, we will identify the triple
scattering paths for which configuration averaging of
$\vec{x}^{\;(4)}$ and $\vec{y}^{\;(4)}$ yields the non-vanishing
contributions around the backwards direction ${\bf k}=-{\bf k}_L$.

\subsection{Selection of triple scattering paths}
\label{sec:preselect}
\subsubsection{Interaction matrices and amplitudes of excitation tranfer}
\label{sec:inter_matr}
In the framework of the master equation approach, multiple scattering paths
can be defined according to the physical meaning of the interaction matrix $V$ as describing
the excitation transfer processes between the atoms \cite{shatokhin10}.
\begin{figure}
\includegraphics[width=4cm]{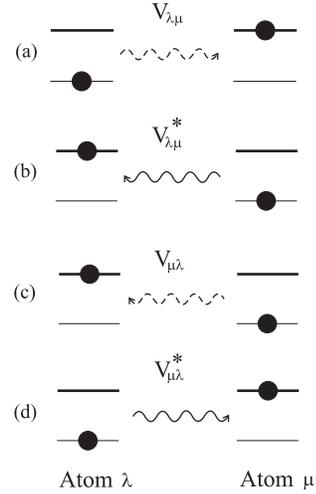}
\caption{Elementary excitation transfer processes between atoms
$\lambda$ and $\mu$ generated by the dipole-dipole interaction
matrix $V$. Black spots correspond to the atomic states after the
exchange process has occurred, and solid (dashed) arrows -- to
positive (negative) frequency amplitudes. The positive (negative)
frequency amplitude of a transfer process from atom $\lambda$ to
atom $\mu$ is described by the matrix $V^*_{\mu\lambda}$ (d)
($V_{\lambda\mu}$ (a)), while the reversed process is described by matrix
$V^*_{\lambda\mu}$ (b) ($V_{\mu\lambda}$ (c)).} \label{fig:dipdip}
\end{figure}
This matrix can be decomposed into a sum \be V=\sum_{\lambda\neq
\mu=1}^3(V_{\lambda\mu}+V_{\lambda\mu}^*), \label{matV1} \e with
$V_{\lambda\mu}\propto T_{\lambda\mu}$ and $V^*_{\lambda\mu}\propto
T^*_{\lambda\mu}$. Each term in the right hand side of
Eq.~(\ref{matV1}) has the same block structure as the matrix $V$
itself (that is, given by Eq.~(\ref{eq:matr_A_V})), and generates an amplitude for a particular
excitation transfer process between the atoms $\lambda$ and $\mu$
(see Fig.~\ref{fig:dipdip}). This amplitude corresponds to a
propagation of the positive- (solid arrow) or negative-frequency
(dashed arrow) field between the atoms \cite{shatokhin10}.
Accordingly, the block matrices $(U^*_{\lambda\mu})_\alpha$,
$(W^*_{\lambda\mu})_\alpha$ ($\alpha=_\ur,\llc,\times$) are associated with a
propagation of the positive-, and $(U_{\lambda\mu})_\alpha$,
$(W_{\lambda\mu})_\alpha$ with the negative-frequency amplitude
between the atoms $\lambda$ and $\mu$.

After the disorder averaging, those triple scattering processes which can be
reduced to each other by relabeling the atomic indices result in identical contributions. Therefore, it is
instructive to identify distinct types of triple scattering paths
and consider one representative thereof. The intensity
(\ref{eq:CBS_Int}) then follows after multiplying the expression corresponding to
a given type of triple scattering process by the number of times it
occurs.

Without restricting the generality, subsequently we will be
concerned with the triple scattering processes whose positive
frequency amplitudes include the path ${\bf
r}_1\rightarrow {\bf r}_2$, and are thus
 proportional to $T^*_{21}$. It is easy to see that there are two triple scattering positive-frequency amplitudes containing the scattering process ${\bf
r}_1\rightarrow {\bf r}_2$: ${\bf
r}_1\rightarrow {\bf r}_2\rightarrow {\bf r}_3$  and ${\bf
r}_1\rightarrow {\bf r}_2\leftarrow {\bf r}_3$. The first and second amplitudes are proportional to  $T^*_{21}T^*_{32}$ and $T^*_{21}T^*_{23}$, respectively.

 Because the atomic
 positions are random, the dipole-dipole coupling constants (see
Eq.~(\ref{dipdip}))
 carry random phases $e^{ik_Lr_{\lambda\mu}}$, for which we assume, since $k_Lr_{\lambda\mu}\gg 1$ as stated above, a uniform distribution
between $0$ and $2\pi$. Then, to survive the disorder averaging,
the positive-frequency triple scattering amplitudes must
be combined with corresponding negative-frequency (conjugate) amplitudes. On top of that, the interference terms in
Eq.~(\ref{eq:CBS_Int}) have additional, ${\bf k}$-sensitive random
phases. This imposes further restrictions on the conjugate
amplitudes.

For a given set of four interaction matrices describing a particular
triple scattering process there are $4!=24$ different terms (due to
the non-commutativity of the matrices). Fortunately,
the phase of each of these 24 terms is the same. Therefore, to see
whether a given process survives the configuration averaging it
suffices to consider only one term.

Using the above restrictions and analyzing different triple
scattering processes, we have identified four
non-equivalent contributions which survive the configuration
averaging: two ladder and two crossed contributions. Below, we will
discuss them separately.

\subsubsection{Ladder intensity}
\label{sec:ladder} The ladder intensity is given by the first line
in the right hand side of Eq.~(\ref{eq:CBS_Int}). We established
that either of the two positive-frequency triple scattering processes, ${\bf
r}_1\rightarrow {\bf r}_2\rightarrow {\bf r}_3$  and ${\bf
r}_1\rightarrow {\bf r}_2\leftarrow {\bf r}_3$,
can be combined with one negative-frequency triple scattering process, to yield a phase-independent contribution.

For the first excitation transfer process, proportional to $T_{21}^*T_{32}^*$ and generated by
matrices $V_{21}^*$, $V_{32}^*$, the conjugate amplitude is generated by matrices
$V_{12}$, $V_{23}$, and leads to a phase-independent solution
for $\la\s_3^+\s_3^-\ra_{\rm ss}^{(4)}$. The corresponding triple
scattering process which is composed of the co-propagating complex
conjugate amplitudes, is shown in Fig.~\ref{fig:triple1}(a). Here,
atom 3 is the source of the backscattered signal, depicted by one
dashed and one solid arrows which are associated with the operators
$\s^+$ and $\s^-$, respectively.

By simply exchanging the atomic indices 1, 2, 3, it is easy to see that the total number of triple scattering
processes which yield the equivalent expressions upon the disorder
averaging is equal to $3!=6$.

Analyzing the second type of the triple scattering process which is generated by matrices $V_{21}^*$, $V_{23}^*$, we identified that, the only conjugate process which leads to a contribution to the ladder intensity is generated by matrices  $V_{12}$ and $V_{32}$. This triple scattering process results in a phase-independent solution for $\la \s_2^+\s_2^-\ra^{(4)}$, and is depicted in Fig.~\ref{fig:triple1}(b). Since the corresponding term is symmetric with respect to a permutation between the indices of the outside atoms, there are in total 3 contributions of this type.

Combining the contributions of both types, which we call ``ladder type 1" and ``ladder type 2" in the following, we obtain the following expression for the total ladder intensity
\be
L^{(3)}_{\rm tot}=L^{(3)}_{{\rm tot},1}+L^{(3)}_{{\rm tot},2},\label{eq:L^3_tot}
\e
with
\be
L^{(3)}_{{\rm tot},1}=6\la\la\s_3^+\s_3^-\ra^{(4)}\ra_{\rm conf},\ 
L^{(3)}_{{\rm tot},2}=3\la\la\s_2^+\s_2^-\ra^{(4)}\ra_{\rm conf},\label{eq:L^3_tot2}
\e
where, from now on, we will for brevity drop the subscript `ss' in
quantum mechanical steady state averages. Furthermore, the total ladder intensity splits into an elastic and an inelastic component:
\be L^{(3)}_{{\rm tot},1}=L^{(3)}_{{\rm el},1}+L^{(3)}_{{\rm inel},1},\ L^{(3)}_{{\rm tot},2}=L^{(3)}_{{\rm el},2}+L^{(3)}_{{\rm inel},2},
\e
where the elastic component results from factorizing the expectation value in Eq.~(\ref{eq:L^3_tot2}), i.e.,
\be
L^{(3)}_{{\rm el},1}=6\la\la\s_3^+\ra\la\s_3^-\ra^{(4)}\ra_{\rm conf},
\e
and similarly for $L^{(3)}_{{\rm el},2}$. The non-factorizable inelastic remainder
 $L^{(3)}_{{\rm inel},i}=L^{(3)}_{{\rm tot},i}-L^{(3)}_{{\rm el},i}$, $i=1,2$, finally, is obtained as an integral over the frequency $\nu$ of the detected photon:
 \be
 L^{(3)}_{{\rm inel},i}=\int_{-\infty}^\infty d\nu L^{(3)}_{{\rm inel},i}(\nu),\ i=1,2,
 \e
where the spectrum $L^{(3)}_{{\rm inel},i}(\nu)$ results from the Fourier transform of the non-factorizable part of the corresponding atomic correlation function
$\langle\sigma^+(t+\tau)\sigma^-(t)\rangle-\langle\sigma^+(t+\tau)\rangle\langle\sigma^-(t)\rangle$ occuring in
Eq.~(\ref{eq:L^3_tot2}) with respect to $\tau$.

\subsubsection{Crossed intensity}
\label{sec:crossed} Let us start discussing the crossed intensity by considering the processes $\propto T^*_{21}T^*_{32}$.
 In accordance with our interpretation of the excitation transfer processes and their
association with the atomic raising and lowering operators, atom
 3 now emits a positive frequency amplitude contributing to the CBS signal (solid arrow).
 Correspondingly,
 the negative frequency amplitudes (dashed arrows) of the outgoing field can be emitted either
 by atoms 1 or 2. Therefore, there are two types of two-atom correlation
 functions,
$\la\s_1^+\otimes\s_3^-\ra^{(4)}$ and
$\la\s_2^+\otimes\s_3^-\ra^{(4)}$, which contribute to the
interference intensity, and will be called ``crossed type 1" and ``crossed type 2" in the following.

\begin{figure}
\includegraphics[width=7cm]{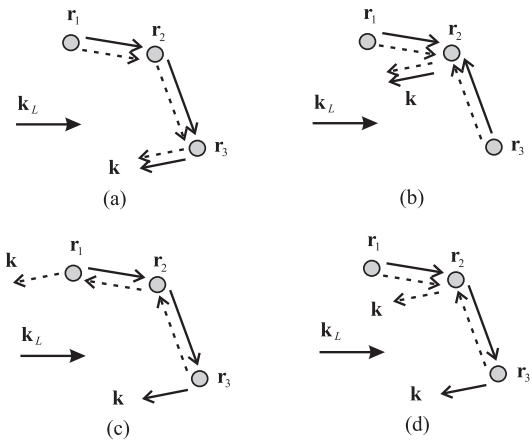}
\caption{Triple scattering processes surviving disorder averaging. Co-propagating amplitudes in diagrams (a) and (b), called ``ladder type 1'' and ``ladder type 2'', in the following, have no
overall phase and contribute to the background intensity; (c)
interference of counter-propagating amplitudes between atoms 1 and
3 (``crossed type 1"), and (d) interference of amplitudes resulting from reversal of the
amplitudes connecting atoms 2 and 3, carry phase factors
$\exp\{i({\bf k}+{\bf k}_L)\cdot{\bf r}_{13}\}$ and $\exp\{i({\bf
k}+{\bf k}_L)\cdot{\bf r}_{23}\}$ (``crossed type 2"), respectively (see
Eq.~(\ref{eq:C^3_tot}) and preceding discussion), and thus contribute only in backscattering direction ${\bf k}=-{\bf k}_L$ where these phases vanish.}
\label{fig:triple1}
\end{figure}

As follows from Eq.~(\ref{eq:CBS_Int}), the atomic dipole
correlation function $\la\s_1^+\otimes\s_3^-\ra^{(4)}$ (type 1) contributes
to the CBS signal upon the configuration averaging if it includes
the position-dependent phase factor which cancels itself with the
phase $\exp(i{\bf k}\cdot{\bf r}_{13})$. We found that this happens only for the counter-propagating amplitudes which are proportional to
$T^*_{21}T^*_{32}T_{21}T_{32}$ (Fig.~\ref{fig:triple1}(c)). In this
case, the function $\la\s_1^+\otimes\s_3^-\ra^{(4)}$ carries the
phase $\exp(i{\bf k}_L\cdot{\bf r}_{13})$ and, hence, the
interference term survives the configuration averaging in the
exact backscattering direction (that is, for ${\bf k}=-{\bf k}_L$
\cite{ftn1}). A contribution of such
counter-propagating amplitudes to the backscattered signal is generic for the CBS effect \cite{sheng}.

The term
$\la\s_2^+\otimes\s_3^-\ra^{(4)}$ (type 2) leads to another type of
crossed contribution. It survives the disorder averaging in the
backwards direction for the scattering process $\propto
T^*_{21}T^*_{32}T_{12}T_{32}$. For this contribution, the negative frequency
amplitude connecting atoms 2 and 3 is reversed (see
Fig.~\ref{fig:triple1}(d)), whereas the one connecting atoms 1 and 2 is
co-propagating with the positive-frequency amplitude.
Such a type of interference processes is characteristic also for CBS from nonlinear classical scatterers \cite{wellens06,wellens06b}.

Last but no least, there is an interference contribution whose positive-frequency amplitude is proportional to $T_{21}^*T_{23}^*$.  The diagram for the corresponding triple scattering process which survives the disorder average can be obtained from the diagram in Fig.~\ref{fig:triple1}(d) by substituting each solid arrow by a dashed one and vice versa. As the corresponding correlation function, $\la \s_3^+\otimes \s_2^-\ra^{(4)}$, is the complex conjugate of the expression corresponding to a diagram in Fig.~\ref{fig:triple1}(d), its contribution can be accounted for by taking twice the real part of the expression describing Fig.~\ref{fig:triple1}(d).

Finally, by noting that the degeneracy of each of the crossed contributions is equal to $3!=6$, we can write
\be
C^{(3)}_{\rm tot}=C^{(3)}_{{\rm tot},1}+C^{(3)}_{{\rm tot},2},\label{eq:C^3_tot}
\e
with
\begin{align}
C^{(3)}_{{\rm tot},1}=&6\la\la\s_1^+\otimes\s_3^-\ra^{(4)}e^{i{\bf k}\cdot{\bf r}_{13}}
\ra_{\rm conf},\label{eq:C^3_tot2}\\
C^{(3)}_{{\rm tot},2}=&12\re [\la\la\s_2^+\otimes\s_3^-\ra^{(4)}e^{i{\bf k}\cdot{\bf
r}_{23}}\ra_{\rm conf}]\label{eq:C^3_tot3}.
\end{align}
Like in the ladder case, also the crossed intensity splits into an elastic and inelastic component
$C^{(3)}_{{\rm tot},i}=C^{(3)}_{{\rm el},i}+C^{(3)}_{{\rm inel},i}$, $i=1,2$, defined as the factorizable and non-factorizable components of the atomic correlation function.

\section{Diagrammatic representation of the triple scattering contribution}
\label{sec:results}
Our goal is to obtain analytical expressions for the
triple scattering ladder and crossed intensities defined in the previous Sec.~\ref{sec:preselect},
and to present them as compounds of single-atom spectral response functions. Since the derivation of these single-atom response functions starting from the master equation amounts to a quite lengthy calculation, we will, for the sake of clarity, first introduce the diagrammatic representation of the  single-atom response functions and define the rules according to which these single-atom building blocks are connected with each other, before showing that the thereby diagrammatically obtained triple scattering signals are equivalent to the expressions derived from the master equation.

\subsection{Triple vs. double scattering}

For this purpose, let us recall the case of double scattering. Here, a diagrammatic representation in terms of single-atom building blocks has already been introduced 
\cite{wellens10,geiger10,shatokhin12} and been proven to be equivalent with the double scattering signal derived from the master equation \cite{shatokhin10}. We will therefore use this insight in order to construct analogous diagrams for the case of triple scattering.

\label{sec:multiple-single}
\subsubsection{Multiple scattering as a combination of single-atom responses}

\begin{figure}
\includegraphics[width=6cm]{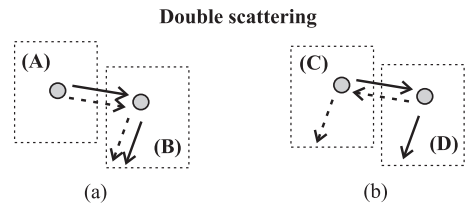}
\caption{Single-atom blocks contributing to double scattering. (a)
Background and (b) interference contributions. Dashed frames
subdivide multiple scattering contributions into single-atom
building blocks.} \label{fig:double1}
\end{figure}

\begin{figure}
\includegraphics[width=7cm]{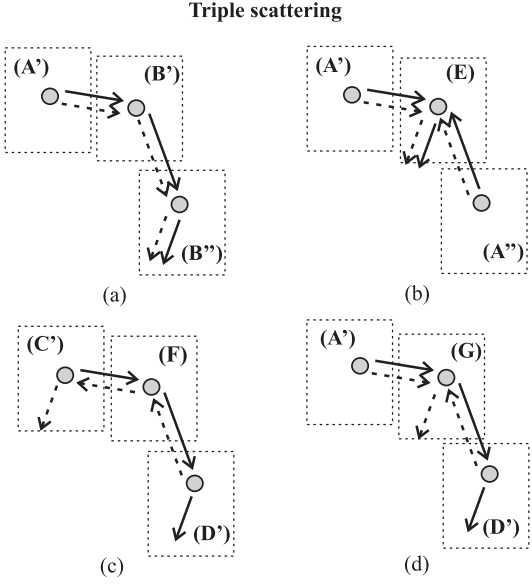}
\caption{Single-atom blocks contributing to triple scattering. (a)
and (b) background; (c) and (d) interference contributions
(corresponding to Fig.~\ref{fig:triple1}(a), (b), (c), and (d),
respectively). Dashed frames subdivide multiple scattering
contributions into single-atom building blocks.} \label{fig:triple}
\end{figure}

To start with, we plot the fundamental processes
which survive the disorder average for the double and triple
scattering contributions in Figs.~\ref{fig:double1} and
\ref{fig:triple}, respectively.

Let us consider the diagrams in Fig.~\ref{fig:double1} which depict
the configuration averaged (a) background and (b) interference
contributions for double scattering. Here and henceforth the laser
field is not displayed, in order to lighten the diagrams. As shown in
\cite{wellens10,geiger10}, both the ladder and crossed spectra can
be obtained by combining single-atom building blocks which are
enclosed in dashed frames (A)-(D) in Fig.~\ref{fig:double1}(a), (b).
The interaction between the atoms is accounted for perturbatively
via the fictitious classical probe fields (solid and dashed arrows)
which are connecting the atoms. These fictitious fields represent
the far-field dipole-dipole interaction between the atoms
\cite{shatokhin10}. Building block (A) is of the zeroth order,
blocks (C), (D) of first, and block (B) of second order in the probe
field or, equivalently, the dipole-dipole coupling strength.
Figure~\ref{fig:triple} shows the composition of the configuration
averaged triple scattering diagrams using
single-atom building blocks enclosed in dashed frames. Comparing the
dashed frames in Figs.~\ref{fig:double1} and \ref{fig:triple} we
impose the compatibility conditions: (A')=(A'')=(A), (B')=(B'')=(B),
(C')=(C), and (D')=(D).

This simple analysis allows us to fully specify the structure of the
contribution to the ladder spectrum described by
Fig.~\ref{fig:triple}(a). It can be compounded from the building
blocks (A) and (B) which can be found by solving the optical Bloch
equations (OBE) under a classical {\it bichromatic} driving
\cite{wellens10,geiger10}.

The second ladder contribution, Fig.~\ref{fig:triple}(b), as well as
the crossed diagrams, Fig.~\ref{fig:triple}(c) and (d), contain
three new building blocks, (E), (F), and (G), respectively. These
blocks describe the spectral response of the intermediate atom,
which receives probe fields from its two neighbors. Therefore,
within the pump-probe approach the blocks (E), (F), and (G) can be
determined by solving the OBE under a classical {\it trichromatic}
driving field.

Because the blocks (A), (B), (C), and (D) familiar from the treatment of double
scattering are important also for triple scattering, we will next recall
 the corresponding single-atom spectral responses.

\subsubsection{Diagrammatic pump-probe approach for double scattering}
\label{sec:pump_probe}

\begin{figure}
\includegraphics[width=8cm]{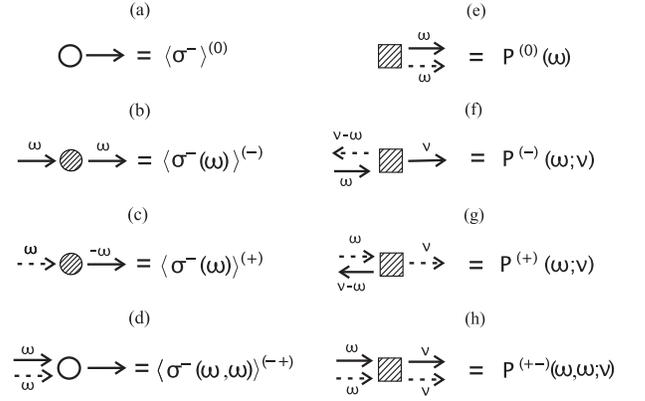}
\caption{Full set of elementary single-atom building blocks for
the double scattering contribution. Circles are (hatched) blank if the
frequency of the outgoing arrow is (different from) equal to the laser frequency.
(a) Complex scattering amplitudes
expressed through the induced atomic dipole moment and (b)-(d) corrections
thereof due to the incoming probe fields; (e)-(f) Building blocks associated with the atomic dipole
correlation functions describing inelastic scattering. In the response
functions on the right hand side, the superscripts $(0)$, $(+)$,
$(-)$ and $(+-)$ or $(-+)$ correspond to the number and type (solid
or dashed) of incoming arrows.} \label{fig:double}
\end{figure}

It is convenient to represent the solutions of the OBE under a
bichromatic driving from which the single-atom building blocks (A),
(B), (C), and (D) (see Fig.~\ref{fig:double1}) are deduced
diagrammatically \cite{shatokhin12}, as displayed in
Fig.~\ref{fig:double}.

Any elastic and inelastic spectral response function describing the
building blocks (A)-(D) can be compounded from circles and boxes
shown in Fig.~\ref{fig:double}. Circles are associated with
expectation values of the atomic dipole moment, and boxes with
Laplace transforms of the steady-state atomic
dipole temporal correlation functions \cite{shatokhin12}. Both types of the averages are
evaluated perturbatively to second order in the weak probe field,
which is represented by the incoming arrows in
Fig.~\ref{fig:double}. The corresponding mathematical expressions are given in Sec.~\ref{sec:expgen}.

Hatching the boxes and some of the circles indicates that the
outgoing arrows have non-zero detunings from the laser frequency.
This helps us to identify the frequencies of the outgoing arrows in
the composition of the double scattering diagrams shown in
Fig.~\ref{fig:double1}(a),(b) -- by virtue of the physical
interpretation of the elementary building blocks as the effective
nonlinear susceptibilities \cite{shatokhin12} which was partially
inspired by the earlier work of Mollow \cite{mollow69,mollow72}.

In all diagrams in Fig.~\ref{fig:double}, the important information about the
incoming and outgoing arrows is encoded in their type (dashed or solid),
direction (incoming or outgoing), and frequency (more precisely, the detuning
from the laser frequency). By convention, we do not label the arrows
that are on-resonant with the laser frequency.
The notation of the spectral response functions on the right hand side
of each diagram in Fig.~\ref{fig:double} embodies the relevant information
about the shapes and arrows of the elementary blocks, and is explained in detail in Sec.~\ref{sec:expgen}.

In general, the diagrams in Fig.~\ref{fig:double} correspond to the
complex-valued spectral response functions. Their conjugate diagrams
are obtained by replacements of each solid arrow by a dashed one and
vice versa. To obtain the corresponding spectral response functions,
we must flip all signs in the symbolic definitions thereof. For
instance,
$(\la\s^-(\omega,\omega)\ra^{(-+)})^*=\la\s^+(\omega,\omega)\ra^{(+-)}
=\la\s^+(\omega,\omega)\ra^{(-+)}$, where the latter equality
follows because the incoming arrows are frequency-degenerate. By the
same token, $(P^{(+)}(\omega;\nu))^*=P^{(-)}(\omega;\nu)$, whereas
$P^{(+-)}(\omega,\omega;\nu)=P^{(-+)}(\omega,\omega;\nu)$ is a real
function.

How to compose the building blocks (A)-(D) from the elementary diagrams shown
in Fig.~\ref{fig:double} was discussed in Ref.~\cite{shatokhin12}. The same setting will be generalized below in Sec.~\ref{sec:results} to describe spectral response functions with arbitrary number of the probe fields.

\subsection{Ladder contribution}
\subsubsection{Type 1}
\label{sec:ladder_spectrum} 
\begin{figure}
\includegraphics[width=7cm]{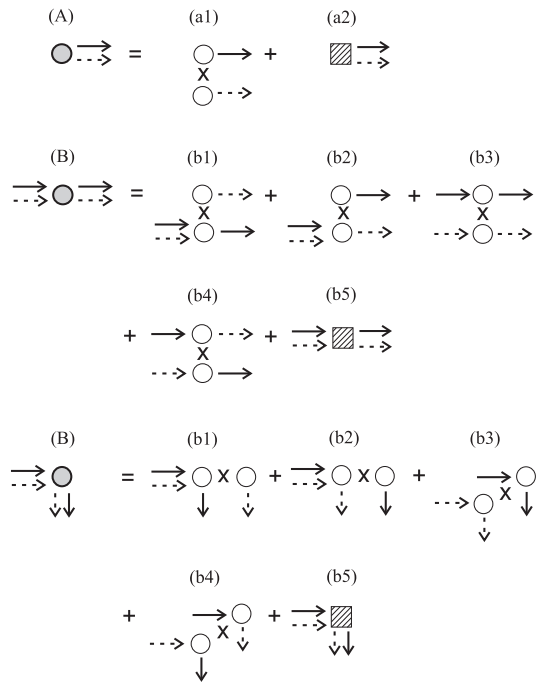}
\caption{(Left) single-atom building blocks contributing to the
ladder spectrum of type 1; (right) expansion of the single-atom
building blocks in elementary single-atom building blocks (see
Fig.~\ref{fig:double}). Circles and boxes describe the elastic and
inelastic responses, respectively. The backscattered signal is
depicted by the downward arrows.} \label{fig:expladd}
\end{figure}

The terms yielding the ladder spectrum
of type 1 are shown in Fig.~\ref{fig:expladd} in the form of three
graphical equations. The left hand sides of these equations depict
the single-atom building blocks (A), (B), (B) from
Fig.~\ref{fig:triple}(a). To allow for an easy identification of the
fields contributing to the detected signal, here and henceforth the
arrows representing the backscattered fields are directed downwards.
Apart from this distinction, which becomes important when combining
the single-atom responses into triple scattering contributions, the
two building blocks (B) in Fig.~\ref{fig:expladd} are equivalent.

The right hand sides of the graphical equations from
Fig.~\ref{fig:expladd} are the expansions of the building blocks
into sums of the elastic and inelastic spectral response functions
shown in Fig.~\ref{fig:double}. The arrows are not labeled by the
frequency values in Fig.~\ref{fig:expladd}, because these values are
uniquely defined at the stage of self-consistent combination of the
single-atom responses into triple scattering diagrams (see below
Sec.~\ref{sec:self-cons}).

Each of the elastic responses is depicted as a product (denoted by
`$\times$') of two elementary building blocks from
Fig.~\ref{fig:double}(a-d) (as well as their complex conjugates).
There are $2^n$ such products -- the number of ways in which $n$
incoming arrows can be distributed among the two circles. Last but
not least, every building block contains one elementary inelastic
building block depicted by a hatched box. The boxes (a2) and (b5) in
Fig.~\ref{fig:expladd} correspond to diagrams of
Fig.~\ref{fig:double}(e) and (h), respectively .

Thus, the block (A) is equal to a sum of 2 terms, whereas the block
(B) is equal to a sum of 5 terms. Consequently, a compound (A)(B)(B)
yields the expression for the ladder intensity $L^{(3)}_{{\rm tot},1}$ of type 1, see Eq.~(\ref{eq:L^3_tot2}), consisting
of 50 terms. Note that, as discussed in Sec.~\ref{sec:ladder}, see Eq.~(\ref{eq:L^3_tot2}), the sum of these terms must be multiplied with a factor 6 in order to account for all different possibilities of exchanging the three atoms with each other.

\subsubsection{Type 2}
\begin{figure}
\includegraphics[width=7cm]{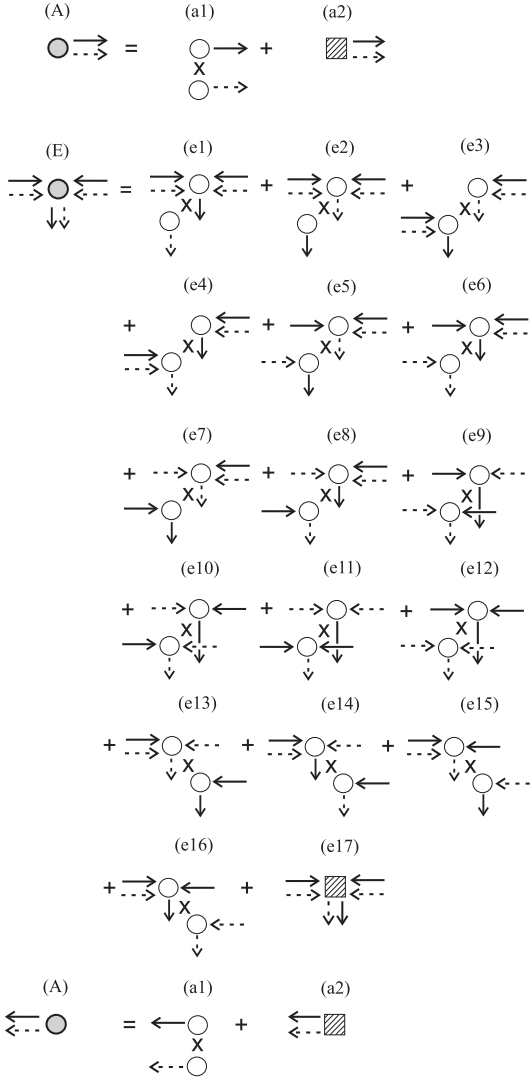}
\caption{(Left) single-atom building blocks contributing to the
ladder spectrum of type 2; (right) expansion of the single-atom
building blocks in elementary single-atom building blocks.}
\label{fig:expladd2}
\end{figure}
To obtain this contribution, we perform a diagrammatic expansion of
Fig.~\ref{fig:triple}(b) into elementary single-atom building
blocks, as shown in Fig.~\ref{fig:expladd2}. Since the middle atom
receives four arrows, the decomposition of the elastic response
functions consists of $2^4=16$ terms, see Fig.~\ref{fig:expladd2}(e1-e16).

\begin{figure}
\includegraphics[width=8.5cm]{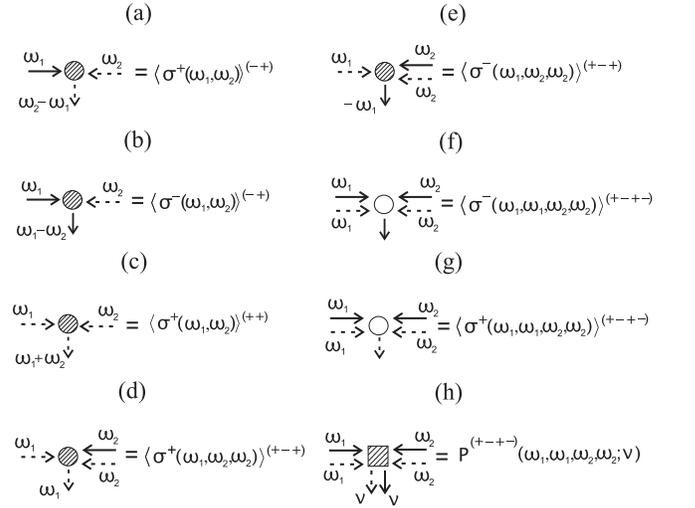}
\caption{Elementary building blocks required to describe the
response of the intermediate atom in Fig.~\ref{fig:expladd2} account
for 2, 3, and 4 arrows with frequencies $\omega_1$ and $\omega_2$.
(a)-(g) scattering amplitudes; (h) frequency correlation function.}
\label{fig:addladd2}
\end{figure}

Obviously, some of the elementary blocks in Fig.~\ref{fig:expladd2}
describing the response of the middle atom, in particular, those
with more than $2$ incoming arrows and some with $2$ incoming arrows, are not
present in Fig.~\ref{fig:double}. Therefore, we depict in Fig.~\ref{fig:addladd2} the new
spectral response functions required for describing the
building block (E). All elementary blocks in Fig.~\ref{fig:addladd2} contain 
incoming arrows which originate from different atoms. Therefore, in general, the values of the incoming frequencies $\omega_1$ and $\omega_2$ are not equal to each other. In the framework of the pump-probe approach, the evaluation of the corresponding spectral response functions requires solving single-atom OBE under a classical trichromatic driving. In the degenerate case 
$\omega_1=\omega_2=\omega$, the functions
$\la\s^\pm(\omega_1,\omega_2)\ra^{(-+)}$ are reduced to the functions 
$\la\s^\pm(\omega,\omega)\ra^{(-+)}$ familiar from double scattering (see Fig.~\ref{fig:double}).

Now, all elementary building blocks in
Fig.~\ref{fig:expladd2} can be expressed in terms of the
elementary building blocks shown in Figs.~\ref{fig:double} and
\ref{fig:addladd2} or in terms of their complex conjugate diagrams. Again, the mathematical expressions for the 
elementary building blocks shown in
Fig.~\ref{fig:addladd2} (as well as those shown in Fig.~\ref{fig:addtriple1}, see Sec.~\ref{sec:crossed1}) are given in Sec.~\ref{sec:expgen}, see Eqs.~(\ref{eq:expvecS1}) and (\ref{eq:P}).
 
By combining all
diagrams (A), (E), (A), we obtain $68=2\times17\times 2$ terms, the sum of which must finally be multiplied by a factor 3 in order to obtain the expression $L^{(3)}_{{\rm tot},2}$ for the ladder intensity of type 2, see Eq.~(\ref{eq:L^3_tot2}).

\subsection{Crossed contribution}
\label{sec:crossed1}
\subsubsection{Type 1}

\begin{figure}
\includegraphics[width=7cm]{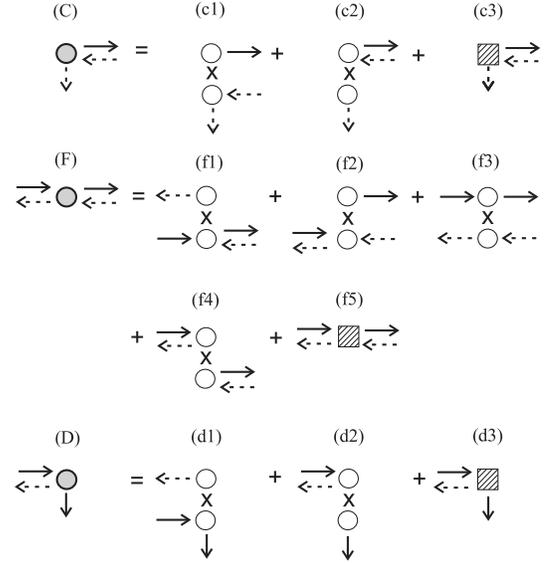}
\caption{(Left) single-atom building blocks for the crossed
contribution of type 1 arising due to interference between the counter-propagating
amplitudes, see Fig.~\ref{fig:triple}(c). The detected signal
originates from the blocks (C) and (D); (right) expansion of the right
hand side into the elementary single-atom building blocks.}
\label{fig:expcross1}
\end{figure}

Recall that this contribution describes interference between
counter-propagating amplitudes. Its graphical representation shown
in Fig.~\ref{fig:expcross1} is constructed analogously to the ladder
terms
by expanding, on the right hand side, each of the single-atom building blocks
(C), (F), and (D)
into a sum of elastic and inelastic responses.

\begin{figure}
\includegraphics[width=8.5cm]{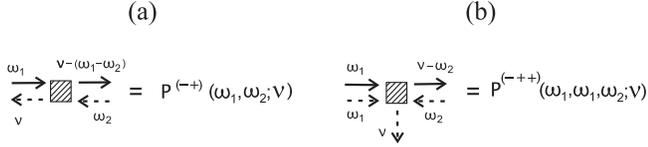}
\caption{Elementary building blocks required to describe the inelastic response of 
the intermediate atom: (a) block (f5) in Fig.~\ref{fig:expcross1}; (b) block (g9) in Fig.~\ref{fig:expcross2}.} \label{fig:addtriple1}
\end{figure}

The diagrammatic expansion of the block (F), apparently, contains the new response function $P^{(-+)}(\omega_1,\omega_2;\nu)$ shown in 
Fig.~\ref{fig:addtriple1}(a).
However, in full analogy with the functions
$\la\s^\pm(\omega_1,\omega_2)\ra^{(+-)}$, the function
$P^{(-+)}(\omega_1,\omega_2;\nu)$ is the same as the function $P^{(-+)}(\omega,\omega;\nu)$ shown
in Fig.~\ref{fig:double}(h), but taken with different frequency
arguments corresponding to the incoming arrows.

Combining all contributions (C), (F), and (D) yields 45 terms.
However, 13 of them contain closed loops (see Appendix \ref{app:forbidden}), and are forbidden according to the rules of
combining diagrams (see Sec.~\ref{sec:self-cons}). Only the remaining 32
triple scattering diagrams (multiplied by a factor 6) thus contribute to the crossed intensity $C^{(3)}_{{\rm tot},1}$ of type 1, see Eq.~(\ref{eq:C^3_tot2}).

\subsubsection{Type 2}
\begin{figure}
\includegraphics[width=7cm]{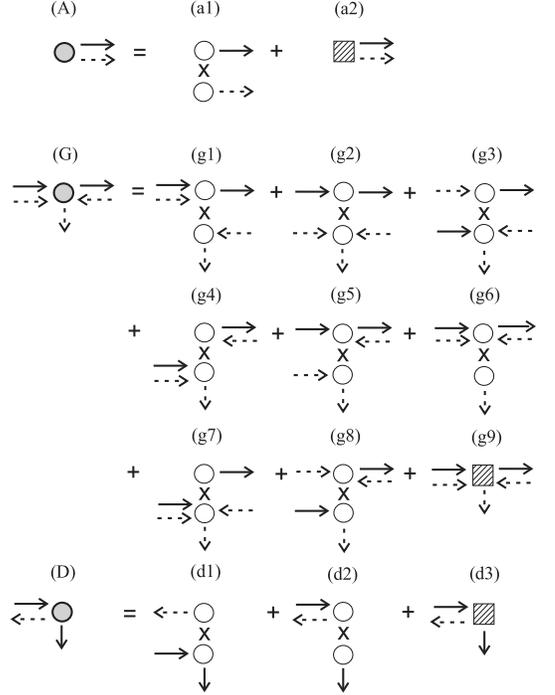}
\caption{(Left) single-atom building blocks for the crossed
contribution of type 2 arising due to interference between the counter-propagating
amplitudes, see Fig.~\ref{fig:triple}(d). The detected signal
originates from the blocks (G) and (D); (right) expansion of the right
hand side into the elementary single-atom building blocks.} \label{fig:expcross2}
\end{figure}

Expanding the single-atom blocks (A), (G), and (D) into the elastic and inelastic
responses, we obtain 2, 9, and 3 diagrams, respectively, in the right hand side of
Fig.~\ref{fig:expcross2}. Analyzing the diagrams contributing to block (G),
we notice a new term displayed in Fig.~\ref{fig:addtriple1}(b).

Combining single-atom responses on the right hand side of
Fig.~\ref{fig:expcross2} yields 56 terms. Excluding the diagrams featuring closed loops
(8 diagrams, see Appendix \ref{app:forbidden}) we come up with 46 terms. Finally, the crossed 
 intensity $C^{(3)}_{{\rm tot},2}$ of type 2 results as 12 times the real part, see Eq.~(\ref{eq:C^3_tot3}), of the sum of all these 46 terms.

\subsection{Self-consistent combination of single-atom building blocks}
\label{sec:self-cons} Figures \ref{fig:expladd}, \ref{fig:expladd2},
\ref{fig:expcross1} and \ref{fig:expcross2} present diagrammatic
expansions of the single-atom responses in the elementary
single-atom building blocks. To obtain the triple scattering ladder
and crossed spectra from these expansions, we combine the elementary blocks
in a self-consistent way \cite{shatokhin12}.

First, we compose the three-atom diagrams using single-atom blocks
by connecting the outgoing arrows with incoming ones regarding the
direction and character (solid or dashed) thereof. Second, according
to the spectral responses functions associated with the blocks given
in Figs.~\ref{fig:double}, \ref{fig:addladd2}, and
\ref{fig:addtriple1}, we ascribe frequency values to each of the
inelastic (that is, distinct from the laser frequency) arrows in the
resulting diagrams. In doing this, we also keep in mind that the two
outgoing (downward) arrows correspond to the spectral intensity of
the backscattered light at a given frequency, and must have the same
frequency value $\nu$. Finally, all intermediate inelastic frequencies
which alter upon scattering are integrated over.

We already mentioned that 13 diagrams resulting from
Fig.~\ref{fig:expcross1} and 8 -- from Fig.~\ref{fig:expcross2} are
forbidden because they contain closed loops wherein the scattered
amplitudes travel between the atoms without going out (see an
example of a loop in Appendix \ref{app:forbidden}). We note that
similar forbidden diagrams appear also in a nonlinear transport
theory of classical scatters \cite{wellens08}.

\begin{figure}
\includegraphics[width=2.5cm]{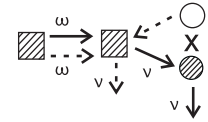}
\caption{Example of a triple scattering interference spectral
response obtained by self-consistently combining diagrams (a2),
(g9), and (d1) from Fig.~\ref{fig:expcross2}.} \label{fig:a2-e9-d1}
\end{figure}

It is instructive to illustrate the
above rules by an example. For that, consider a self-consistent
combination of diagrams (a2)(g9)(d1) (see Fig.~\ref{fig:expcross2}), which constitutes 1 of the 46 diagrams contributing to the crossed intensity of type 2. The corresponding compound
diagram is drawn in Fig.~\ref{fig:a2-e9-d1}.
The symbolic expressions for the response functions corresponding to
the elementary building blocks appearing in Fig.~\ref{fig:a2-e9-d1}
are given in Figs.~\ref{fig:double}(a), (b), (e), and
\ref{fig:addtriple1}(b) (more precisely, the block in
Fig.~\ref{fig:double}(a) is the complex conjugate of the one from
Fig.~\ref{fig:a2-e9-d1}). Noting that the intermediate inelastic
frequency $\omega$ changes its value, we obtain the following result
for the contribution: \begin{align}
\text{(a2)(g9)(d1)}&=\int_{-\infty}^{\infty}d\omega
P^{(0)}(\omega)P^{(-++)}(\omega,\omega,0;\nu)\n\\
&\times\la\s^+\ra^{(0)}\la\s^-(\nu)\ra^{(-)},\label{eq:(a2)(e9)(d1)}\end{align}
where we have omitted the prefactor $\la T_{21}^*T^*_{32}T_{12}T_{32}\ra_{\rm conf}$ which scales as $(k_L\ell)^{-4}$ (see Eq.~(\ref{dipdip})), where $\ell$ is the mean interatomic distance.
Any other combination of the elementary building blocks is expressed in an analogous way.
Further examples are considered in Sec.~\ref{sec:example}.
\begin{figure}
\includegraphics[width=4.5cm]{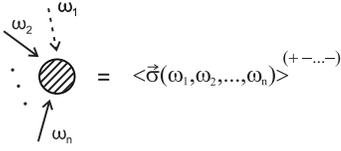}
\caption{
Circle with incoming dashed and solid arrows at
frequencies $\omega_1,\omega_2, \ldots, \omega_n$, representing the $n$th
order correction to the optical Bloch vector
$\la\vs(\omega_1,\omega_2,\ldots,\omega_n)\ra^{(+-\ldots -)}$, whose
frequencies and signs in the superscript are arranged in the same
order. The signs `$+$' and `$-$' are associated with dashed and
solid arrows, respectively. The positive and negative frequency amplitudes
associated with this block have frequency $\sum_i s_i\omega_i$ which is in general
non-zero, i.e. detuned from the laser frequency, and hence the circle is hatched.} \label{fig:bloch_vector_n_arrows}
\end{figure}

\section{General expressions for single-atom building blocks}
\label{sec:expgen} In the previous Secs.~\ref{sec:solving_meq} and
\ref{sec:results} we introduced diagrams for all elementary single
atom building blocks, see Figs.~\ref{fig:expladd}, \ref{fig:expladd2}, \ref{fig:expcross1}
and \ref{fig:expcross2}, exhibiting at most four incoming arrows.
In this section, we will give the corresponding mathematical expressions,
generalized to an arbitrary number $n$ of incoming probe fields.

It is important to have
formulae for an arbitrary $n$ for the following reasons: First, the laser-driven cold atoms in
a cloud experience random polychromatic fields which can be modeled
by a large number of the incident arrows labeled by different
frequencies \cite{geiger10}. Second, already in case of three
incoming arrows the expressions for the building blocks become
rather heavy. It is these two considerations that urged us to look
for a general and concise description of the building blocks. Such a
formulation is possible due to the clear hierarchical
structure of the building blocks and the small number of constituents
needed for their construction.

To compose an arbitrary spectral response function, we need (i) the single-atom Green's function $G(z)=(z-M)^{-1}$, where $M=M_\lambda$ according to Eq.~(\ref{eq:BlochM}) with $\Omega_\lambda=\Omega$ for an atom placed at the coordinate origin; (ii) the matrices $\Delta^{(+)}$ and $\Delta^{(-)}$ describing the coupling of the negative- and positive-frequency probe fields to the atomic dipole (Eq.~(\ref{eq:defDmDp})); (iii) frequency values and types (dashed or solid) of all the incoming probe fields.

\subsection{Spectral response functions associated with the Bloch vector}
\label{sec:bbel}
In the hierarchy of the single-atom building blocks, the
most basic element is the steady state optical Bloch vector for an atom
interacting with a laser field alone. We will represent this vector
graphically by a blank
circle: \be \bigcirc=\la\vs\ra^{(0)}=G\vec{L}, \e where $\vec{L}$ is given in Eq.~(\ref{eq:BlochM}), and $G\equiv G(0)$.

Circles with one, two, ..., $n$ incoming arrows
 describe perturbative
corrections to the Bloch vector  due to one, two,..., $n$
weak classical probe fields. Equivalently, these corrections can be
regarded as arising from the excitation transfer processes from
the surrounding atoms (see Sec.~\ref{sec:inter_matr}). A graphical representation
of the $n$th order correction
$\la\vs(\omega_1,\omega_2,\ldots,\omega_n)\ra^{(+-\ldots -)}$
for a particular choice of the incoming arrows is shown in Fig.~\ref{fig:bloch_vector_n_arrows}.

As follows from Fig.~\ref{fig:bloch_vector_n_arrows}, we associate
signs $s_j$ ($j=1,\ldots,n$) with each of the $n$ incoming arrows
according to the rule \be s_j=\begin{cases}
-, \text{for incoming}\; \longrightarrow , \\
+, \text{for incoming} \;\dasharrow. 
\end{cases}
\e We stress that, to avoid confusion, we always arrange frequencies
and the corresponding signs in the same order.

With these preliminary remarks, we introduce the general expression
for the vector
$\la\vs(\omega_1,\omega_2,\ldots,\omega_n)\ra^{(s_1s_2\ldots s_n)}$:
\begin{widetext}
\be \la\vs(\omega_1,\omega_2,\ldots,\omega_n)\ra^{(s_1 s_2\ldots
s_n)}= \sum_{\pi(j_1,\ldots,j_n)} G(i\mathsmaller\sum_{k=1}^n
s_{j_k}\omega_{j_k})\Delta^{(s_{j_n})}\ldots
 G(is_{j_1}\omega_{j_1}+is_{j_2}\omega_{j_2})
 \Delta^{(s_{j_2})}
G(is_{j_1}\omega_{j_1}) \Delta^{(s_{j_1})}G\vec{L},
\label{eq:expvecS1} \e
\end{widetext}
where $\pi(j_1,\ldots,j_n)$ denotes $n!$ permutations of indices
$j_1, \ldots, j_n\in \{1,\ldots,n\}$. As can be proven by the method of induction,
the same expression is obtained when expanding the quasi-stationary solution of the optical Bloch equations for an atom driven by a superposition of the laser field
plus $n$ additional fields with different frequencies $\omega_1,\ldots,\omega_n$ up to
first order in each of the additional field's amplitudes.

To illustrate application of the formula (\ref{eq:expvecS1}) by
an example, we will consider the function
$\la\vs(\omega_1,\omega_2)\ra^{(+-)}$ describing a second-order
correction to the Bloch vector due to the incoming negative and
positive frequency amplitudes at frequencies $\omega_1$ and
$\omega_2$, respectively (Fig.~\ref{fig:addladd2}(b)). By virtue of Eq.~(\ref{eq:expvecS1}) we
obtain:
\begin{align}
\la\vs(\omega_1,&\omega_2)\ra^{(+-)}=G(i[\omega_1-\omega_2])\Dm
G(i\omega_1)\Dp
G\vec{L}\n\\
&+G(i[\omega_1-\omega_2])\Dp G(-i\omega_2)\Dm G\vec{L}.
\label{eq:ds2}
\end{align}

The first and second element of the vector
$\la\vs(\omega_1,\ldots,\omega_n)\ra^{(s_1 \ldots s_n)}$,
$\la\s^-(\omega_1,\ldots,\omega_n)\ra^{(s_1 \ldots s_n)}$ and
$\la\s^+(\omega_1,\ldots,\omega_n)\ra^{(s_1 \ldots s_n)}$,
correspond to the outgoing positive and negative frequency
amplitude depicted by a solid and dashed arrow, respectively.
The simplest outgoing amplitudes correspond to absent incoming arrows:
\begin{align}
\bigcirc\!\!\longrightarrow &=\la\s^-\ra^{(0)}=\bigl[G\vec{L}\,\bigr]_1,\\
\bigcirc\!\!\dasharrow&=\la\s^+\ra^{(0)}=\bigl[G\vec{L}\,\bigr]_2.
\end{align}
The frequencies of the outgoing arrows are uniquely defined by those
of the incoming ones and are equal to $-\sum_j s_j\omega_j$ for a
solid and $+\sum_j s_j\omega_j$ for a dashed arrow. Unless, for each
arrow with non-zero detuning $\omega_j$ there exists a corresponding complex conjugate arrow
with the same detuning, $\sum_j s_j\omega_j\neq 0$. In Fig.~\ref{fig:bloch_vector_n_arrows}, we have assumed that the latter is true, and therefore hatched the circle.

\subsection{Spectral response functions associated with
correlation functions}
\label{sec:bbinel}

Previously, we described the amplitude response functions
which are depicted graphically by a circle with $n\geq 0$ incoming and
a single
outgoing arrows. Now, we will consider the
building blocks with two outgoing arrows, of which one
is solid and one dashed.
In general, these building blocks are associated with the
atomic dipole correlation functions \cite{shatokhin12}.
As already mentioned above, splitting these functions into a factorizable and non-factorizable component then defines the elastic and inelastic component of the radiated intensity.

\subsubsection{Factorized response functions} In the beginning, we
will consider the  building blocks
which can be factorized as products of the negative and positive frequency scattering
amplitudes. In the simplest case of no incoming arrows, such a
product gives the intensity of the elastic component of resonance
fluorescence. Graphically, this quantity is represented as
\begin{align}
\bigcirc  \!&\!\dasharrow \n\\
\times& \\
\bigcirc \!&\!\longrightarrow.\n
\end{align}
For an arbitrary number $n$ of the incoming arrows,
they can be distributed among the outgoing positive and negative
frequency amplitudes of a factorized correlation function in $2^n$
ways. The resulting correlation function is then expanded into a sum of
these $2^n$ combinations. For $n\leq 3$, the diagrammatic expansions
are given in Figs.~\ref{fig:expladd}, \ref{fig:expcross1}, and
\ref{fig:expcross2}.

In the following, we will use permutations between the sets of indices
$\{j_1,\ldots, j_k\}$ and
$\{j_{k+1},\ldots,j_n\}$, which we
will denote by $\pi(j_1,\ldots, j_k|j_{k+1},\ldots,j_n)$.
There are $n!/k!(n-k)!$ such permutations.
Now, if there are $n$ incoming arrows, the factorized response
function $g^{(s_1\ldots s_n)}(\omega_1,\ldots,\omega_n)$ reads
\begin{widetext}
\begin{align} g^{(s_1\ldots
s_n)}(\omega_1,\ldots,\omega_n)&=
\la\s^+(\omega_1,\ldots,\omega_n)\ra^{(s_1\ldots s_n)}\la\s^-\ra^{(0)}\n\\
&+\sum_{\pi(j_1,\ldots,j_{n-1}|j_n)}\la\s^+(\omega_{j_1},\ldots,\omega_{j_{n-1}})\ra^{(s_{j_1}\ldots s_{j_{n-1}})}\la\s^-(\omega_{j_n})\ra^{(s_{j_n})}\n\\
&+\ldots\n\\
&+\sum_{\pi(j_1|j_2,\ldots,j_n)}\la\s^+(\omega_{j_1})\ra^{(s_{j_1})}\la\s^-(\omega_{j_2},\ldots,\omega_{j_n})\ra^
{(s_{j_2}\ldots
s_{j_n})}\n\\
&+\la\s^+\ra^{(0)}\la\s^-(\omega_1,\ldots,\omega_n)\ra^{(s_1\ldots
s_n)}. \label{eq:Pel}
\end{align}
Let us illustrate
Eq.~(\ref{eq:Pel}) by an example for $n=2$:
\begin{align}
g^{(+-)}(\omega_1,\omega_2)&=\la\s^+(\omega_1,\omega_2)\ra^{(+-)}\la\s^-\ra^{(0)}
+\la\s^-(\omega_1,\omega_2)\ra^{(+-)}\la\s^+\ra^{(0)}\n\\
&+\la\s^+(\omega_1)\ra^{(+)}\la\s^-(\omega_2)\ra^{(-)}
+\la\s^-(\omega_1)\ra^{(+)}\la\s^+(\omega_2)\ra^{(-)}, \label{eq:g2}
\end{align}
\end{widetext}
where the subsequent terms of Eq.~(\ref{eq:g2}) correspond to diagrams (f1), (f2), (f3), and (f4) with the frequencies of the incoming dashed and solid arrows equal to $\omega_1$ and $\omega_2$, respectively (see Fig.~\ref{fig:expcross1}).

\subsubsection{Non-factorized response functions}
\begin{figure}
\includegraphics[width=6cm]{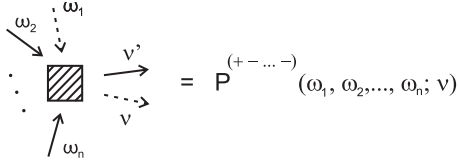}
\caption{Inelastic building blocks have two outgoing arrows. The frequencies of the
outgoing and incoming arrows are related by the energy conservation relation
  $\nu-\nu^\prime=\mathsmaller\sum_k s_k\omega_k$ (see Eq.~(\ref{eq:ener_cons})).} \label{fig:box_N_arrows}
\end{figure}

The fluctuating part of the atomic dipole correlation function cannot be
factorized. The building blocks associated with such functions are depicted by
hatched boxes with one dashed and one solid outgoing arrows (see
Fig.~\ref{fig:box_N_arrows}), and are denoted by $P^{(s_1 \ldots s_n)}(\omega_1,\ldots,\omega_n;\nu)$.
In this case, the frequencies of the outgoing arrows are not defined
only by those of the incoming ones: an additional frequency $\nu$ appears which describes the spectral distribution of the outgoing, say dashed, arrow. The frequency $\nu'$ of the outgoing solid arrow is related to $\nu$ by the identity
\be \nu'\equiv \nu-\sum_{k=1}^ns_{k}\omega_{k},
\label{eq:ener_cons} \e
which reflects the energy conservation upon the scattering process.

It is well known how to obtain the inelastic part of the emission spectrum of a laser-driven atom (see, for instance, \cite{cohen-tannoudji}).
To find the function $P^{(s_1 \ldots s_n)}(\omega_1,\ldots,\omega_n;\nu)$, we have generalized the same procedure to account, in addition to the laser field, for the presence of $n$ probe fields.

An important ingredient in the definition of the function $P^{(s_1 \ldots s_n)}(\omega_1,\ldots,\omega_n;\nu)$ are the expressions for the fluctuating part
of the $n$th order correction of atomic dipole correlation functions.
These expressions can be straightforwardly obtained with the aid of
the $n$th-order correction to the Bloch vector $\la\vs(\omega_1,\ldots,\omega_n)^{(s_1\ldots s_n)}$, Eq.~(\ref{eq:expvecS1}), and the factorized response function $g^{(s_1\ldots s_n)}(\omega_1,\ldots,\omega_n)$, Eq.~(\ref{eq:Pel}).

Using the function $g^{(s_1\ldots s_n)}(\omega_1,\ldots,\omega_n)$, we create two vector functions, $\vec{g}_{+}^{\;(s_1\ldots
s_n)}(\omega_1,\ldots,\omega_n)$ and $\vec{g}_{-}^{\;(s_1\ldots
s_n)}(\omega_1,\ldots,\omega_n)$. The former and latter vectors are obtained
by making substitutions $\s^+\rightarrow \vs $ and $\s^-\rightarrow \vs $, respectively, in each of the terms of Eq.~(\ref{eq:Pel}). Explicitly,
\begin{widetext}
\begin{align} \vec{g}_{\pm}^{\;(s_1\ldots
s_n)}(\omega_1,\ldots,\omega_n)&=
\la\vs(\omega_1,\ldots,\omega_n)\ra^{(s_1\ldots s_n)}\la\s^\mp\ra^{(0)}\n\\
&+\sum_{\pi(j_1,\ldots,j_{n-1}|j_n)}\la\vs(\omega_{j_1},\ldots,\omega_{j_{n-1}})\ra^{(s_{j_1}\ldots s_{j_{n-1}})}\la\s^\mp(\omega_{j_n})\ra^{(s_{j_n})}\n\\
&+\ldots\n\\
&+\sum_{\pi(j_1|j_2,\ldots,j_n)}\la\vs(\omega_{j_1})\ra^{(s_{j_1})}\la\s^\mp(\omega_{j_2},\ldots,\omega_{j_n})\ra^
{(s_{j_2}\ldots s_{j_n})}\n\\
&+\la\vs\ra^{(0)}\la\s^\mp(\omega_1,\ldots,\omega_n)\ra^{(s_1\ldots
s_n)}. \label{eq:vecg}
\end{align}
Now, the fluctuating part of the atomic dipole correlation function reads:
\be
\vec{q}_{\pm}^{\;(s_1\ldots s_n)}(\omega_1,\ldots,\omega_n)=\mp i\Delta^{(\pm)}
\la\vs(\omega_1,\ldots,\omega_n)\ra^{(s_1\ldots s_n)}
-\vec{g}_\pm^{\;(s_1\ldots s_n)}(\omega_1,\ldots,\omega_n).
\label{eq:q_pm}
\e
\end{widetext}
In case of no incoming arrows, Eq.~(\ref{eq:q_pm}) reduces to
the following expressions:
\begin{align*}
\vec{q}_+^{\;(0)}&=-i\Dp \la\vs\ra^{(0)}+\vec{L}_1-\la\vs\ra^{(0)}\la\s^-\ra^{(0)},\\
\vec{q}_-^{\;(0)}&=i\Dm \la\vs\ra^{(0)}+\vec{L}_2-\la\vs\ra^{(0)}\la\s^+\ra^{(0)},
\end{align*}
where the $\vec{L}_1=(0,1/2,0)^T$, and $\vec{L}_2=(1/2,0,0)^T$ appear due to the identity $\la\s^+\s^-\ra^{(0)}\equiv(1+\la\s^z\ra^{(0)})/2$.

Vectors $\vec{q}_{\pm}^{\;(s_1\ldots s_k)}(\omega_1,\ldots,\omega_k)$ ($k=0,1,\ldots, n$) enter the definition of the inelastic building block shown in Fig.~\ref{fig:box_N_arrows} through the relations:
\begin{widetext}
\be
P^{(s_1\ldots s_n)}(\omega_1,\ldots,\omega_n;\nu)=\frac{1}{2\pi}\lt(P_+^{(s_1\ldots s_n)}(\omega_1,\ldots,\omega_n;\nu)+P_-^{(s_1\ldots s_n)}(\omega_1,\ldots,\omega_n;\nu)\rt),
\label{eq:P}
\e
where
\beml
\begin{align}
P_+^{(s_1\ldots s_n)}(\omega_1,\ldots,\omega_n;\nu)&= \sum_{\pi(j_1,\ldots,j_n)}
\lt[G(i\nu)\Delta^{(s_{j_n})}\ldots
 \Delta^{(s_{j_2})}G(i\nu'+is_{j_1}\omega_{j_1})
\Delta^{(s_{j_1})}G(i\nu')\vec{q}_+^{\;(0)}\rt]_2\n\\
&+\sum_{\pi(j_2,\ldots,j_n)}\lt[G(i\nu)\Delta^{(s_{j_n})}\ldots
 \Delta^{(s_{j_2})}G(i\nu'+is_{j_1}\omega_{j_1})\vec{q}_+^{\;(s_{j_1})}(\omega_{j_1})\rt]_2\n\\
&+\ldots \n\\
&+\lt[G(i\nu)\vec{q}_+^{\;(s_{j_1}\ldots
s_{j_n})}(\omega_{j_1},\ldots,\omega_{j_n})\rt]_2,\\
P_-^{(s_{1}\ldots s_n)}(\omega_1,\ldots,\omega_n;\nu)&= \sum_{\pi(j_1,\ldots,j_n)} \lt[G(-i\nu')\Delta^{(s_{j_n})}\ldots
 \Delta^{(s_{j_2})} G(-i\nu+is_{j_1}\omega_{j_1})
 \Delta^{(s_{j_1})}G(-i\nu)\vec{q}_-^{\;(0)}\rt]_1\n\\
&+\sum_{\pi(j_2,\ldots,j_n)}\lt[G(-i\nu')\Delta^{(s_{j_n})}\ldots\Delta^{(s_{j_2})}G(-i\nu+is_{j_1}\omega_{j_1})
\vec{q}_-^{\;(s_{j_1})}(\omega_{j_1})\rt]_1\n\\
&+\ldots \n\\
&+\lt[G(-i\nu')\vec{q}_-^{\;(s_{j_1}\ldots
s_{j_n})}(\omega_{j_1},\ldots,\omega_{j_n})\rt]_1,
\end{align}
\label{P+P-}
\eml
\end{widetext}
with $\nu^\prime$ defined in Eq.~(\ref{eq:ener_cons}). Again, as can be proven by the method of induction, Eqs.~(\ref{eq:P}-\ref{P+P-}) coincide with the expressions obtained from the perturbative solution for a single atom driven by a laser field and additional weak probe fields.

Finally, we will illustrate Eqs.~(\ref{eq:P}), (\ref{P+P-}) by the
simplest non-trivial example of the expression for a hatched box
with one incoming dashed arrow at frequency $\omega$ (see, e.g.,
Fig.~\ref{fig:expcross1}(c3)). In this case we have \beq
P^{(+)}(\omega;\nu)&=& \frac{1}{2\pi}\lt(\lt[G(i\nu)\Dp G(i\nu-i\omega)\vec{q}_+^{\;(0)}\rt]_2\rt.\n\\
&+&\lt.\lt[G(i\nu)\vec{q}_+^{\;(+)}(\omega)\rt]_2\rt.\n\\
&+& \lt.\lt[ G(i\omega-i\nu)\Dp G(-i\nu)\vec{q}_-^{\;(0)}\rt]_1\rt.\n\\
&+&\lt. \lt[ G(i\omega-i\nu)\vec{q}_-^{\;(+)}(\omega)\rt]_1\rt). \eq

\subsection{Derivation from the master equation}
\label{sec:outline_derive}

Having defined the elementary single-atom building blocks in Secs.~\ref{sec:bbel} and \ref{sec:bbinel}, and the rules for connecting them with each other in
Sec.~\ref{sec:results}, we are now able to calculate all the various components (ladder and crossed of type 1 and 2) of the triple scattering intensity in terms of single-atom quantities. It remains to be shown that the corresponding expressions are identical to those derived from the master equation introduced in Sec.~\ref{sec:solving_meq}.

To obtain the solution of the master equation in the form of self-consistent
combination of single-atom building blocks, we generalize the method
that was previously
applied to two atoms \cite{shatokhin10}. Some steps towards such a solution have been already made
in Sec.~\ref{sec:solving_meq}.
Namely, in
Sec.~\ref{sec:recurr} we introduced the recurrence relations for
vectors $\vec{x}^{\;(n)}$, $\vec{y}^{\;(n)}$, and $\vec{z}^{\;(n)}$.
Thereafter, in Sec.~\ref{sec:preselect}, we selected the triple
scattering diagrams for which the incoherent and interference
intensities survive the disorder average, resulting in the ladder
and crossed contributions.

The next step is to factorize all terms containing the relevant
vectors $\vec{x}^{\;(4)}$, $\vec{y}^{\;(4)}$ as combinations of
three single-atom expressions. Recall that these vectors include
four interaction matrices sandwiched between the Green's functions
$G_+$, $G_\times$, or $D_\times$ (see Eq.~(\ref{eq:recurr})). The
factorization is then accomplished by using the integral
representations for $G_\times$ and $D_\times$ which express the two-
and three-atom evolution matrices as frequency integrals over tensor
products of the single-atom Green's matrices (see Appendix
\ref{ap-Dx}).

After this procedure, each term contributing to the ladder and
crossed intensity contains multiple (up to sevenfold) frequency
integrals. Although each integrand represents a product of three
single-atom expressions, neither of them can be interpreted
physically. To obtain the physically meaningful single-atom building blocks as defined by the solution of the optical Bloch equations for a single atom,
we simplify these expressions further.

To this end, by using the sum rules (see Appendix
\ref{sec:sum_rules}) we solve exactly some of the frequency
integrals. As a result, we obtain the final expressions for the
ladder and crossed spectra which include at most two integrations
(see, e.g., Eq.~(\ref{eq:asymspect}) valid for very strong driving
$\Omega\gg \gamma$). The ladder spectrum (Fig.~\ref{fig:triple}(a)
and (b)) is given by a sum of $118=50+68$ terms corresponding to all
possible combinations of the single-atom building blocks shown in
Fig.~\ref{fig:expladd} ($50=2\times5\times 5$) and
Fig.~\ref{fig:expladd2} ($68=2\times 17\times 2$), see
Sec.~\ref{sec:ladder_spectrum} above. Likewise, the crossed spectrum
(Fig.~\ref{fig:triple}(c),(d)) is given by all allowed combinations
of the single-atom building blocks shown in
Figs.~\ref{fig:expcross1} and \ref{fig:expcross2}, see
Sec.~\ref{sec:crossed1}. This completes the derivation of the triple scattering contribution expressed in terms of single-atom response functions.

\section{Triple scattering spectra}
\label{sec:example} In this section, we present our results for
the triple scattering spectra of CBS. In the beginning, we
provide analytical expressions for the elastic and inelastic
contributions in the limit of small Rabi frequency $\Omega\ll
\gamma$. Thereafter, we consider the non-perturbative regime of strong atom-laser interaction.

\subsection{Case of small Rabi frequencies ($\Omega\ll
\gamma$)} \label{sec:small_Rabi} In this limit, it is natural to
restrict ourselves to the triple scattering contributions that are
not larger than $\sim (\Omega/\gamma)^4$. Physically, this
corresponds to a weakly inelastic scattering by three atoms of the
laser field containing not more than two photons.

Analyzing the behavior of the triple scattering diagrams, we
established that they exhibit different asymptotics when $\Omega\ll
\gamma$. For example, the crossed diagram (a2)(g9)(d1) (see
Eq.~(\ref{eq:(a2)(e9)(d1)})) yields an expression which is $\sim
(\Omega/\gamma)^6$, and can therefore be neglected.
We will consider separately the elastic and inelastic
components of CBS.

\begin{figure}
\includegraphics[width=7cm]{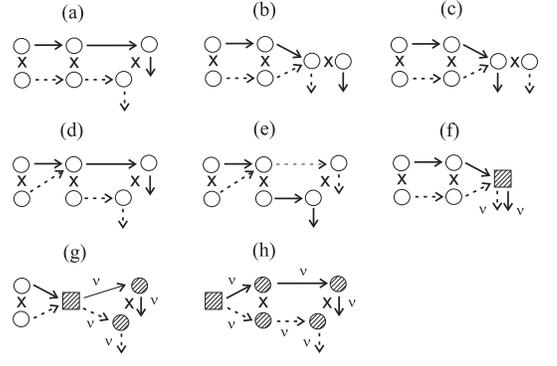}
\caption{Triple scattering diagrams yielding the elastic (a-e) and
inelastic (f-h) ladder spectra of type 1.
For one-photon processes ($\sim \Omega^2$), only
diagram (a) contributes.} \label{fig:ladder_2photon}
\end{figure}
\subsubsection{Elastic spectrum}
Diagrams in Fig.~\ref{fig:ladder_2photon}(a)-(e) describe the
elastic ladder intensity $L^{(3)}_{\rm el,1}$. They can be presented
in the form of equations: (a)=(a1)(b3)(b3), (b)=(a1)(b3)(b2),
(c)=(a1)(b3)(b1), (d)=(a1)(b1)(b3), and (e)=(a1)(b2)(b3), where the
compounds on the right hand sides are the single-atom
building blocks from Fig.~\ref{fig:expladd}.
\begin{figure}
\includegraphics[width=7cm]{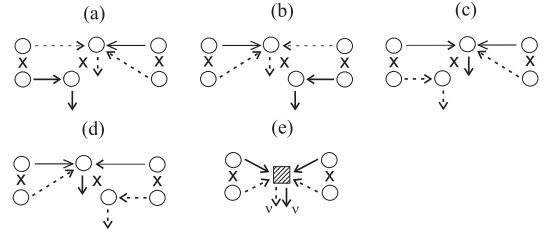}
\caption{Triple scattering diagrams yielding the elastic
(a-d) and inelastic (e) ladder spectra of type 2.}
\label{fig:ladder_2photon2}
\end{figure}
Figure \ref{fig:ladder_2photon2}(a-d) depicts diagrams contributing
to $L^{(3)}_{\rm el,2}$. In this case, the corresponding equations for diagrams read (see Fig.~\ref{fig:expladd2}):
(a)=(a1)(e7)(a1), (b)=(a1)(e13)(a1), (c)=(a1)(e6)(a1), (d)=(a1)(e16)(a1).

\begin{figure}
\includegraphics[width=7cm]{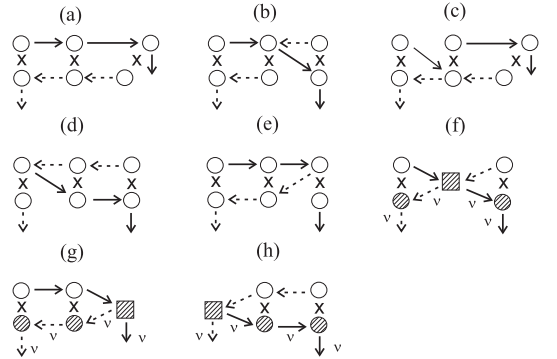}
\caption{Triple scattering diagrams yielding the elastic
(a-e) and inelastic (f-h) crossed spectra of type 1. For one-photon processes, only
diagram (a) which is a reciprocal of diagram (a) in
Fig.~\ref{fig:ladder_2photon}, contributes.}
\label{fig:cross_2photon1}
\end{figure}

\begin{figure}
\includegraphics[width=7cm]{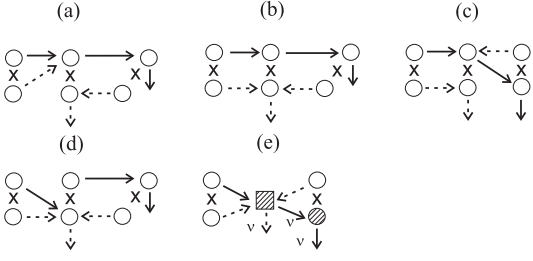}
\caption{Triple scattering diagrams yielding the elastic
(a-d) and inelastic (e) crossed spectra of type 2.}
\label{fig:cross_2photon2}
\end{figure}

Likewise, the elastic crossed contributions in
Fig.~\ref{fig:cross_2photon1}(a)-(e) result from the combinations of
single-atom blocks of Fig.~\ref{fig:expcross1}: (a)=(c1)(f3)(d1),
(b)=(c1)(f1)(d1), (c)=(c1)(f2)(d1), (d)=(c2)(f3)(d1),
(e)=(c1)(f3)(d2), which yield the elastic crossed intensity $C^{(3)}_{\rm el, 1}$.
Finally, the four diagrams in Fig.~\ref{fig:cross_2photon2}
yield the crossed contribution $C^{(3)}_{\rm el, 2}$. They are composed of the
diagrams of Fig.~\ref{fig:expcross2} as follows: (a)=(a1)(g1)(d1),
(b)=(a1)(g2)(d1), (c)=(a1)(g5)(d1), (d)=(a1)(g7)(d1).

We obtain the following analytical expressions for the elastic
triple scattering ladder and crossed intensities which are valid up
to $(\Omega/\gamma)^4$:
\beml
\begin{align}
L^{(3)}_{\rm el, 1}&=\frac{3\tilde{\Omega}^2}{32(1+\tilde{\delta}^2)^3}-
\frac{27\tilde{\Omega}^4}{32(1+\tilde{\delta}^2)^4},\label{eq:Lel1}\\
L^{(3)}_{\rm el, 2}&=-
\frac{6\tilde{\Omega}^4}{32(1+\tilde{\delta}^2)^4},\label{eq:Lel2}\\
 C^{(3)}_{\rm el, 1}&=\frac{3\tilde{\Omega}^2}{32(1+\tilde{\delta}^2)^3}-
 \frac{24\tilde{\Omega}^4}{32(1+\tilde{\delta}^2)^4},\label{eq:cro_el1}\\
C^{(3)}_{\rm el,
2}&=-\frac{24\tilde{\Omega}^4}{32(1+\tilde{\delta}^2)^4},\label{eq:cro2_el}
\end{align}
\label{eq:LelCel}
\eml
where $\tilde{\Omega}= \Omega/\gamma$, $\tilde{\delta}= \delta/\gamma$.
Using the above results, which, as we have checked, coincide with the results following from the
two-photon scattering theory \cite{wellens06}, we draw two conclusions. First, in the linear
regime, only the diagrams in Figs.~\ref{fig:ladder_2photon}(a) and
\ref{fig:cross_2photon1}(a) contribute to the signal, and ladder and crossed components
are equal to each other, what indicates perfect phase coherence
in the linear regime. From Eqs.~(\ref{eq:Lel1}) and
(\ref{eq:cro_el1}) it follows that \be L^{(3)}_{\rm el, 1}=C^{(3)}_{\rm
el, 1}=\frac{3\tilde{\Omega}^2}{32(1+\tilde{\delta}^2)^3}. \e

Second, at increased Rabi frequencies, the
interference contribution $C^{(3)}_{\rm el,
1}+C^{(3)}_{\rm el,
2}$ decreases faster than the ladder contribution $L^{(3)}_{\rm
el,1}+L^{(3)}_{\rm
el,2}$, since the contributions proportional to $\tilde{\Omega}^4$ have a negative sign, and are larger for the crossed component ($24+24=48$) than for the ladder component ($27+6=33$), see Eqs. (\ref{eq:LelCel}). The fact that the interference contribution exceeds the background can be explained by the fact that, in the case of nonlinear triple scattering, the coherent backscattering signal is in general formed by {\em three} interfering amplitudes \cite{wellens06}.

\subsubsection{Inelastic spectrum}
\label{sec:small_inelastic} The four inelastic ladder diagrams
are depicted in Fig.~\ref{fig:ladder_2photon}(f)-(h) and Fig.~\ref{fig:ladder_2photon2}(e). These diagrams
result from the
combinations: (f)=(a1)(b3)(b5), (g)=(a1)(b5)(b3), (h)=(a2)(b3)(b3) (see Fig.~\ref{fig:expladd}), (e)=(a1)(e17)(a1) (see Fig.~\ref{fig:expladd2}).

Concerning the inelastic crossed spectrum, it is given by the four
contributions shown in Fig.~\ref{fig:cross_2photon1}(f)-(h) and Fig.~\ref{fig:cross_2photon2}(e). Diagrams
(f)-(h) consist of the following building blocks: (f)=(c1)(f5)(d1),
(g)=(c1)(f3)(d3), and (h)=(c3)(f3)(d1), with the right hand sides
coming from Fig.~\ref{fig:expcross1}. ``Type 2'' crossed diagram in
Fig.~\ref{fig:cross_2photon2}(e) yields one contribution to the
inelastic spectrum in the weakly inelastic regime: (e)=(a1)(g9)(d1),
with the constituents defined in Fig.~\ref{fig:expcross2}. Due to this contribution which emerges when the scatterers are
nonlinear \cite{wellens06,wellens08}, the interference effect
modifies dramatically: the enhancement factor associated with the
inelastic scattering becomes greater than 2, again due to the interference between three amplitudes mentioned above.

Indeed, consider the analytical expressions for the inelastic ladder
and crossed spectra:
\beml
\begin{align}
L^{(3)}_{\rm inel,1}(\tilde{\nu})&=\frac{1}{2\pi}\frac{6\tilde{\Omega}^4}{32(1+\tilde{\delta}^2)^3}\label{eq:Lin}\\
&\times
\frac{(3(1+\tilde{\delta}^2)+4\tilde{\delta}\tilde{\nu}+2\tilde{\nu}^2)^2}
{(1+(\tilde{\delta}-\tilde{\nu})^2)
(1+(\tilde{\delta}+\tilde{\nu})^2)^3}, \n\\
L^{(3)}_{\rm inel,2}(\tilde{\nu})&=\frac{1}{2\pi}\frac{\tilde{\Omega}^4}{32(1+\tilde{\delta}^2)^3}\\
&\times\frac{12}{(1+(\tilde{\delta}-\tilde{\nu})^2)(1+(\tilde{\delta}+\tilde{\nu})^2)}\n\\
 C^{(3)}_{\rm
inel,1}(\tilde{\nu})&=\frac{1}{2\pi}\frac{6\tilde{\Omega}^4}{4(1+\tilde{\delta}^2)^3} \label{eq:c3_1inel}\\
&\times\frac{(1+\tilde{\delta}(\tilde{\delta}+\tilde{\nu}))^2}
{(1+(\tilde{\delta}-\tilde{\nu})^2)
(1+(\tilde{\delta}+\tilde{\nu})^2)^3},\n\\
C^{(3)}_{\rm inel, 2}(\tilde{\nu})&=\frac{1}{2\pi}\frac{6\tilde{\Omega}^4}{4(1+\tilde{\delta}^2)^3}\label{eq:C2_inel}\\
&\times\frac{1+\tilde{\delta}(\tilde{\delta}+\tilde{\nu})}
{(1+(\tilde{\delta}-\tilde{\nu})^2)(1+(\tilde{\delta}+\tilde{\nu})^2)^2}.\n
\end{align}
\label{eq:inelast_small_Rabi}
\eml
\begin{figure}
\includegraphics[width=6cm]{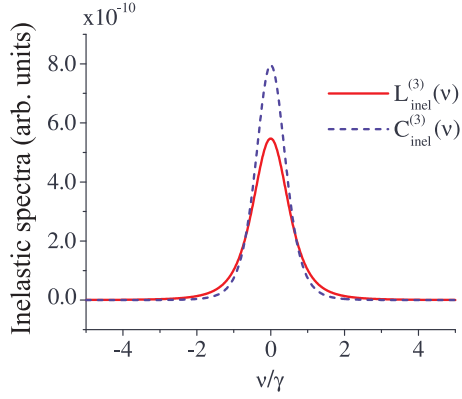}
\caption{(Color online) Inelastic ladder and crossed spectra of
triple scattering for the case of exact resonance ($\delta=0$) and
$\Omega=0.01\gamma$.}\label{fig:0inel_sp_resonant}
\end{figure}
As we have checked, the inelastic spectra given by Eqs.~(\ref{eq:inelast_small_Rabi}) are also
in full agreement with the  two-photon diagrammatic scattering theory
\cite{wellens06}. The curves corresponding to the functions $L^{(3)}_{\rm
inel}(\nu)=L^{(3)}_{\rm inel,1}(\nu)+L^{(3)}_{\rm inel,2}(\nu)$
and $C^{(3)}_{\rm
inel}(\nu)=C^{(3)}_{\rm inel,1}(\nu)+C^{(3)}_{\rm inel,2}(\nu)$, are
plotted in Fig.~\ref{fig:0inel_sp_resonant}.

It is clearly seen that the dashed line corresponding to
$C^{(3)}_{\rm inel}(\nu)$ is above the solid line
corresponding to $L^{(3)}_{\rm inel}(\nu)$. To characterize this
effect in more quantitative terms, let us,
following \cite{wellens06}, assume that the elastic component has
been filtered out, and study the behavior of the enhancement factor,
which in this case is defined as \be \eta=1+\frac{C^{(3)}_{\rm
inel}}{L^{(3)}_{\rm inel}}. \label{eq:enhancement}\e Integrating
Eqs.~(\ref{eq:inelast_small_Rabi})
over $\tilde{\nu}$, we obtain: \begin{align} C^{(3)}_{\rm
inel}&=\frac{3(22+\tilde{\delta}^2)\tilde{\Omega}^4}{128(1+\tilde{\delta}^2)^4},\\
L^{(3)}_{\rm inel}&=\frac{3(154+25\tilde{\delta}^2+3\tilde{\delta}^4)
\tilde{\Omega}^4}{1024(1+\tilde{\delta}^2)^4}, \end{align} wherefrom
we deduce that $\eta(\tilde{\delta})>2$ for
$|\tilde{\delta}|<1.042$, with the maximum value of $\approx
2.143$ at $\tilde{\delta}=0$. In fact, in a
cloud consisting of a large number of atoms, and applying appropriate frequency filtering, the enhancement factor (\ref{eq:enhancement}) can in principle
reach the value 3 \cite{wellens06b}.

So far, we have restricted our consideration to the case of weakly
inelastic scattering which is valid for very small Rabi frequencies.
We will next address the behavior of the CBS spectra in the
non-perturbative regime of the atom-light interaction.

\subsection{Non-perturbative account of atom-laser interaction}
\label{sec:nonpert}
\begin{figure}
\includegraphics[width=8cm]{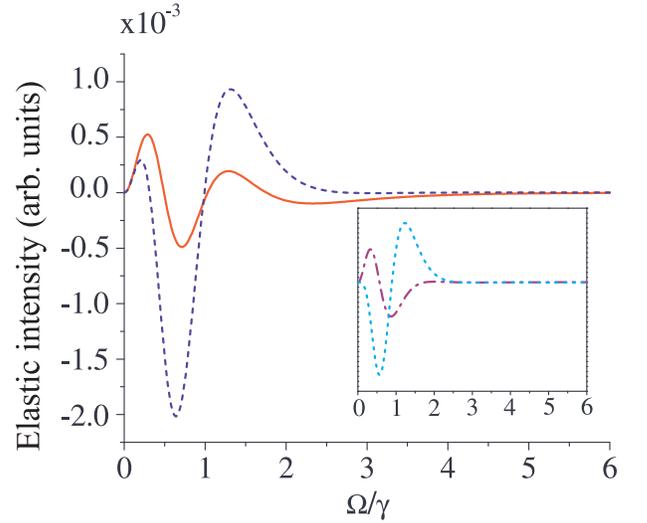}
\caption{(Color online) Elastic intensities as a function of Rabi
frequency at exact resonance: (solid)
 $L^{(3)}_{\rm el}$, (dashed) $C^{(3)}_{\rm el}$. Inset displays
 two elastic crossed contributions:
 (dashed-dotted) $C^{(3)}_{\rm el,1}$; (dotted) $C^{(3)}_{\rm
el,2}$.}\label{fig:el_sp_resonant}
\end{figure}

If the Rabi frequency does not satisfy the condition $\Omega\ll
\gamma$, we enter the regime of light scattering by atoms wherein
the non-perturbative expressions for the single-atom building blocks
must be used. Consequently, we consider contributions from
all triple scattering diagrams resulting from self-consistent
combinations of the single-atom building blocks shown in
Figs.~\ref{fig:expladd}, \ref{fig:expladd2}, \ref{fig:expcross1}, and
\ref{fig:expcross2}. The results of our numerical calculations of
the triple scattering elastic and inelastic spectra are presented in
Figs.~\ref{fig:el_sp_resonant} and \ref{fig:inel_sp_resonant},
respectively.

\subsubsection{Elastic spectrum}
The elastic intensity exhibits an oscillatory-like behavior as a
function of the Rabi frequency for $\Omega\lesssim 2\gamma$.
The intensity can assume
negative values, which is in accordance with the fact that the
triple scattering is observed from the laser-driven transition.
Therefore, it is only a correction to the physical intensity which
also includes single scattering, and is always positive.

A curious feature exhibited by the crossed contributions in
Fig.~\ref{fig:el_sp_resonant}(inset) is that, in a range of Rabi
frequencies where the elastic intensity is significant, they have
opposite phases. For values of $\Omega\lesssim  0.5\gamma$,
$C^{(3)}_{\rm el,2}<0<C^{(3)}_{\rm el,1}$. If $0.5\gamma\lesssim \Omega
\lesssim \gamma$, both crossed contributions feature destructive
interference. In the range $\gamma\lesssim \Omega\lesssim 2\gamma$ the crossed
contributions again exhibit opposite interference character:
$C^{(3)}_{\rm el,1}<0<C^{(3)}_{\rm el,2}$.

Although we have not found a simple explanation of why the two crossed
contributions exhibit opposite interference character, this fact is not surprising
given that they
originate from
different triple scattering paths.

\subsubsection{Inelastic spectrum}
\begin{figure}
\includegraphics[width=8.5cm]{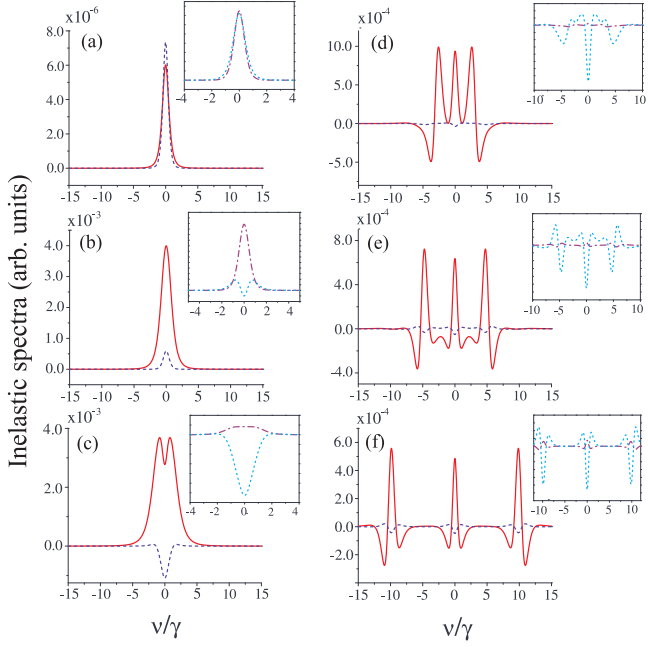}
\caption{(Color online) Inelastic ladder (solid) and crossed
(dashed) spectra of triple scattering for the case of exact
resonance ($\delta=0$) for different Rabi frequencies: $\Omega=$ (a)
$0.1\gamma$; (b) $0.5\gamma$; (c) $\gamma$; (d) $3\gamma$; (e)
$5\gamma$; (f) $10\gamma$. Insets show two inelastic crossed
contributions $C^{(3)}_{\rm inel,1}(\tilde{\nu}) $ (dashed-dotted)
and $C^{(3)}_{\rm inel,2}(\tilde{\nu})$
(dotted).}\label{fig:inel_sp_resonant}
\end{figure}

To see how inelastic processes affect CBS from three atoms, in 
Fig.~\ref{fig:inel_sp_resonant} we plot the ladder and crossed spectra 
for different values of the Rabi frequency. For $\Omega=0.1\gamma$, the two-photon processes still give the dominant contribution to triple scattering, which is manifest in a qualitative agreement between the spectra in Figs.\ref{fig:inel_sp_resonant}(a) and \ref{fig:0inel_sp_resonant}. However, at increased values of the Rabi frequency, Fig.~\ref{fig:inel_sp_resonant}(b-f), the crossed spectra
are dominated by the ladder spectra. Furthermore, small in magnutude
interference spectra feature destructive interference, or CBS anti-enhancement (see Fig.~\ref{fig:inel_sp_resonant}(c-f)), due to the crossed contribution of type 2 (see insets).

Increasing the Rabi frequency is also accompanied by
a splitting of the ladder and crossed spectra into several Lorentzian
and dispersive resonances.  This transformation of the spectra can be
 attributed to the dressing of atomic states by the laser field \cite{shatokhin07}.

For very strong driving strengths
 $\Omega \gg\gamma$, the intensities of the elastic components become
 vanishingly small, and the total spectral line shapes are given by the corresponding
 inelastic spectra. The latter can be deduced by combining the three elementary
 inelastic building blocks (hatched boxes) for each of the contributions. Namely,  the combinations (a2)(b5)(b5),
 (a2)(e17)(a2), (c3)(f5)(d3), and (a2)(g9)(d3) yield $L^{(3)}_{\rm inel,1}(\nu)$, $L^{(3)}_{\rm inel,2}(\nu)$, $C^{(3)}_{\rm inel,1}(\nu)$, and
 $C^{(3)}_{\rm inel,2}(\nu)$, respectively (see Figs.~\ref{fig:expladd}, \ref{fig:expladd2}, \ref{fig:expcross1}, and \ref{fig:expcross2}). The inelastic ladder and crossed spectra
 can then be presented by the compact formulas:
\beml
\label{eq:asymspect}
 \begin{align}
L^{(3)}_{\rm inel,1}(\nu)&=6\int d\omega^\prime d\omega^{\prime\prime}
P^{(0)}(\omega^\prime)\\
&\times P^{(+-)}(\omega^\prime,\omega^\prime;\omega^{\prime\prime})
P^{(+-)}(\omega^{\prime\prime},\omega^{\prime\prime};\nu),\n\\
L^{(3)}_{\rm inel,2}(\nu)&=3\int d\omega^\prime d\omega^{\prime\prime}
P^{(0)}(\omega^\prime)P^{(0)}(\omega^{\prime\prime})\\
&\times P^{(+-+-)}(\omega^\prime,\omega^\prime,\omega^{\prime\prime},\omega^{\prime\prime};\nu),\n\\
C^{(3)}_{\rm inel, 1}(\nu)&=6\int d\omega^\prime
d\omega^{\prime\prime}
P^{(+)}(\omega^\prime;\nu)P^{(-)}(\omega^{\prime\prime};\nu)\n\\
&\times P^{(+-)}(\nu-\omega^{\prime\prime},
\nu-\omega^{\prime};\omega^{\prime})
,\\
C^{(3)}_{\rm inel, 2}(\nu)&=12\re \int d\omega^\prime
d\omega^{\prime\prime}
P^{(0)}(\omega^\prime)P^{(-)}(\omega^{\prime\prime};\nu)\n\\
&\times P^{(+-+)}(\omega^{\prime},\omega^{\prime},
\nu-\omega^{\prime\prime};\nu).
\end{align}
\eml

\section{Discussion and conclusion}
\label{sec:discuss}
We have studied the coherent backscattering of intense laser light from
three two-level atoms
using the master equation approach. The goal of this work was to find the analytical solution of the problem and show its equivalence with the solution following from the diagrammatic pump-probe approach to CBS. We have fulfilled this task in the spirit of the earlier work \cite{shatokhin10}, where technical tools such as the recurrence relations, the integral representations and spectral decompositions of the Green's matrices, as well as the sum rules were developed and applied to prove the equivalence between the master equation and the pump-probe approaches for double scattering. Here, we have generalized these techniques for three atoms.

From the rigorous analytical solution of the master equation in quadratures, we have found the explicit analytical formulas for
the single-atom building blocks whose diagrammatic self-consistent combination
yields the triple scattering ladder and crossed spectral intensities. The obtained expressions for the single-atom spectral response functions are equivalent to the pump-probe solutions following from the optical Bloch equations under classical bi- and trichromatic driving fields.

In contrast to double scattering, triple scattering features two crossed contributions in the inelastic scattering regime which manifest
the multi-wave character of CBS \cite{wellens08}. It is the interplay between these two contributions that defines the behavior of the overall
interference effect as a function of the Rabi frequency. In the weakly inelastic scattering regime, the two inelastic crossed contributions are positive and in phase, yielding the CBS enhancement factor $>2$ -- in qualitative agreement with the earlier prediction for nonlinear classical scatterers \cite{wellens06}. However, increasing the laser
intensity leads to a rapid decrease of the relative intensities of the crossed contributions with respect to the ladder term, on the one hand, and to the destructive interference, on the other hand.

The results of this work go beyond the problem of triple scattering.
We have deduced general
analytical expressions for single-atom building blocks for an arbitrary number of incoming probe fields. This corresponds to the case of multiple scattering of intense laser light in a cloud consisting of  a large number of atoms, where each atom is radiated both by the coherent laser field and by the fields radiated from all other atoms, the sum of which is not necessarily small compared to the laser field. These expressions can therefore be incorporated into the stochastic pump-probe approach \cite{wellens10,geiger10} for an accurate simulation of the multiple inelastic scattering of laser light in cold atomic ensembles, including propagation effects in the effective medium, such as the attenuation of the laser field.

Another direction of future work will be to generalize our present results for realistic atoms probed in the CBS ``saturation'' experiments \cite{chaneliere04,balik05}.
This can be achieved by including the polarization degree of freedom into the description of the single-atom response functions.

\acknowledgements
Enjoyable discussions with Sybille Braungardt and Andreas Ketterer are gratefully acknowledged.
This work was financially supported by DFG through grant BU-1337/9-1.

\appendix
\section{Evolution matrices}
\label{sec:appA}
\subsection{Matrix $A$: evolution of independent atoms}

The blocks of matrix $A$ (see Eq.~(\ref{eq:matr_A_V})) read
\beml
\begin{align}
M_+&=M_3\oplus M_2\oplus M_1,\label{eq:Mp}\\
M_\times &=(M_2\otimes\openone+\openone\otimes M_3)
\oplus(M_1\otimes\openone+\openone\otimes M_3)\n\\
&\oplus (M_1\otimes\openone+\openone\otimes M_2),\label{eq:M_times}\\
L_+&=\lt(\ba{ccc}\vec{L}\otimes\openone&\openone\otimes\vec{ L}&0\\
\vec{ L}\otimes\openone& 0&\openone\otimes\vec{ L}\\
 0&\vec{ L}\otimes\openone&\openone\otimes\vec{ L}\ea\rt),\label{eq:L_plus}\\
L_\times &=\lt(\ba{ccc}
\vec{ L}\otimes\openone\otimes\openone&\openone
\otimes\vec{ L}\otimes\openone&\openone\otimes\openone\otimes\vec{ L}\ea\rt),\label{eq:nullblock}\\
N_\times&=\openone\otimes\openone\otimes M_3+ \openone\otimes
M_2\otimes\openone+M_1\otimes\openone\otimes\openone.\label{eq:N_times}
\end{align}
\label{eq:M1M2L1}
\eml
Here, $\openone$ denotes the unit $3\times 3$ matrix, $M_\lambda$ ($\lambda=1, 2, 3$) is a position-dependent $3\times 3$ matrix describing the evolution of the Bloch vector of atom $\lambda$, and $\vec{L}$ is a three dimensional vector \cite{shatokhin10}. Explicitly,
\be
M_\lambda=\lt(\ba{ccc}-\gamma+i\delta&0&-i\Omega_\lambda/2\\
0&-\gamma-i\delta&i\Omega^*_\lambda/2\\
-i\Omega_\lambda^*&i\Omega_\lambda&-2\gamma\ea\rt),\; \vec{L}=(0,0,-2\gamma)^T.\label{eq:BlochM}\e
To be self-contained, we remind here the form of the optical Bloch equation:\be
\la\dot{\vec{\s}}_\lambda\ra=M_\lambda\la\vec{\s}_\lambda\ra+\vec{L}.
\e
Given the vector $\vec{L}$, we can also define the vector $\vec{\Lambda}$ (see Eq.~(\ref{mateq})):
\be
\vec{\Lambda}=(\vec{ L},\vec{ L},\vec{ L},\vec{0})^T,\label{eq:nullvector}
\e
where the null vector in (\ref{eq:nullvector}) contains 54 zeroes.

Using Eqs.(\ref{eq:Mp}) and (\ref{eq:Greens}) we obtain \be
G_+(z)=G_3(z)\oplus G_2(z)\oplus G_1(z), \label{eq:gplus} \e with
 $G_\lambda(z)=(z-M_\lambda)^{-1}$, cf. Eq.~(\ref{eq:Greens})

\subsection{Matrix $V$: evolution of the dipole-dipole interacting atoms}
\label{eq:matrV} A convenient way to specify the blocks
$(U_{\lambda\mu})_\alpha$, $(W_{\lambda\mu})_\alpha$
($\alpha=_\ur,\llc,\times$) of the matrix $V$, see
Eqs.~(\ref{eq:matr_A_V}) and (\ref{matV1}), is implicit -- by
describing their action on test vectors. This is how it was done in
the case of two atoms \cite{shatokhin10}. In fact, due to the
pairwise nature of the dipole-dipole interaction, the results for
two atoms can, after some modifications, be adapted to the $N$-atom
case. Here, we will sketch this procedure for three atoms.

To this end, we will for the moment abstract ourselves from the
presence of a third atom, and consider how the blocks of the
`contracted' dipole-dipole interaction matrix (denoted as $\bar{v}$) act in the
two-atom
case. When this is done, it is straightforward to define the
action of the blocks of the matrix $V$.

Let $\vec{a}_\lambda$ and $\vec{a}_\mu$ be 3-component column vectors, with the indices $\lambda, \mu$ referring to the atom's number.
Assuming $\lambda<\mu$, we create two test vectors \be
(\vec{p}_{\lambda\mu})_1=(\vec{a}_\mu,\vec{a}_\lambda)^T,\quad(\vec{p}_{\lambda\mu})_2=\vec{a}_\lambda\otimes\vec{a}_\mu.\label{probe_vectors}\e

The `contracted' interaction matrices $\bar{v}_\alpha$ ($\alpha=_\ur,\llc,\times$) are characterized by the
following identities \cite{shatokhin10}: \beml \begin{align}
(\bar{v}_{\lambda \mu})_\ur(\vec{a}_\lambda\otimes\vec{a}_\mu)&=\lt(\ba{c}2iT_{\lambda \mu}\Dp\vec{a}_\mu[\vec{a}_\lambda]_2\\\vec{0}\ea\rt),\label{id-V12-ur}\\
(\bar{v}_{\mu\lambda})_\ur(\vec{a}_\lambda\otimes\vec{a}_\mu)&=\lt(\ba{c}\vec{0}\\2iT_{\mu\lambda}\Dp\vec{a}_\lambda[\vec{a}_\mu]_2\ea\rt),\label{id-V21-ur}\\
(\bar{v}_{\lambda \mu}^*)_\ur(\vec{a}_\lambda\otimes\vec{a}_\mu)&=\lt(\ba{c}\vec{0}\\-2iT_{\lambda \mu}^*\Dm\vec{a}_\lambda[\vec{a}_\mu]_1\ea\rt),\label{id-V12*-ur}\\
(\bar{v}_{\mu\lambda}^*)_\ur(\vec{a}_\lambda\otimes\vec{a}_\mu)&=\lt(\ba{c}-2iT_{\mu\lambda}^*\Dm \vec{a}_\mu[\vec{a}_\lambda]_1\\
\vec{0}\ea\rt),\label{id-V21*-ur}\\
(\bar{v}_{\lambda \mu})_\times(\vec{a}_\lambda\otimes\vec{a}_\mu)&=-2T_{\lambda \mu}\Dm\vec{a}_\lambda\otimes\Dp\vec{a}_\mu,\label{id-V12-t}\\
(\bar{v}_{\mu\lambda})_\times(\vec{a}_\lambda\otimes\vec{a}_\mu)&=-2T_{\mu\lambda}\Dp\vec{a}_\lambda\otimes\Dm\vec{a}_\mu,\label{id-V21-t}\\
(\bar{v}_{\lambda \mu}^*)_\times(\vec{a}_\lambda\otimes\vec{a}_\mu)&=-2T_{\lambda j}^*\Dm\vec{a}_\lambda\otimes\Dp\vec{a}_\mu,\label{id-V12*-t}\\
(\bar{v}_{\mu\lambda}^*)_\times(\vec{a}_\lambda\otimes\vec{a}_\mu)&=-2T_{\mu\lambda}^*\Dp\vec{a}_\lambda\otimes\Dm\vec{a}_\mu,\label{id-V21*-t}\\
(\bar{v}_{\lambda \mu})_\llc\lt(\ba{c}\vec{a}_\mu\\\vec{a}_\lambda\ea\rt)&=\vec{n}_1\otimes(2iT_{\lambda \mu}\Dp\vec{a}_\mu),\label{id-V12-ll}\\
(\bar{v}_{\mu\lambda})_\llc\lt(\ba{c}\vec{a}_\mu\\\vec{a}_\lambda\ea\rt)&=2iT_{\mu\lambda}\Dp\vec{a}_\lambda\otimes\vec{n}_1,\label{id-V21-ll}\\
(\bar{v}_{\lambda \mu}^*)_\llc\lt(\ba{c}\vec{a}_\mu\\\vec{a}_\lambda\ea\rt)&= (-2iT_{\lambda \mu}^*\Dm\vec{a}_\lambda)\otimes\vec{n}_2,\label{id-V12*-ll}\\
(\bar{v}_{\mu\lambda}^*)_\llc\lt(\ba{c}\vec{a}_\mu\\\vec{a}_\lambda\ea\rt)&=\vec{n}_2\otimes(-2iT_{\mu\lambda}^*\Dm\vec{a}_\mu),\label{id-V21*-ll}
\end{align} \label{eq:all_V} \eml where $[\vec{a}_\lambda]_k$ and
$[\vec{a}_\mu]_k$ refer to the $k$-th component ($k=1,2,3$) in the
basis defined by the single-atom Bloch vector
$\la\vs_\lambda\ra=(\la\s_\lambda^-\ra,\la\s_\lambda^+\ra,\la\s_\lambda^z\ra)^T$,
$\vec{n}_1=(\frac{1}{2},0,0)^T$, $\vec{n}_2=(0,\frac{1}{2},0)^T$,
$\vec{0}=(0,0,0)^T$, and \be
\Dm=\lt(\ba{ccc}0&0&-i/2\\0&0&0\\0&i&0\ea\rt),\;\;
\Dp=\lt(\ba{ccc}0&0&0\\0&0&i/2\\-i&0&0\ea\rt).\label{eq:defDmDp} \e
Identities (\ref{eq:all_V}) show that the blocks of the matrix $\bar{v}$ accomplish
mutual mappings between the vectors of the form $\vec{p}_1$ and $\vec{p}_2$ (we omit for the moment the indices $\lambda$, $\mu$):
\be
\bar{v}_\ur: \vec{p}_2\rightarrow \vec{p}_1^{\;\prime}, \;
\bar{v}_\times: \vec{p}_2\rightarrow \vec{p}_2^{\;\prime},\;
\bar{v}_\llc: \vec{p}_1\rightarrow \vec{p}_2^{\;\prime},
\e
where $\vec{p}_j^{\;\prime}$ ($j=1,2$) is the result of the transformation which has the
same structure as the vector $\vec{p}_j$.

Returning to the three-atom case, it can be shown that the matrices
 $U_\alpha$ and $W_\alpha$ ($\alpha=_\ur,\llc,\times$), acting on the
test vectors
\beml
\begin{align}
\vec{P}_1&=(\vec{a}_3,\vec{a}_2,\vec{a}_1)^T,\\
\vec{P}_2&=(\vec{a}_2\otimes\vec{a}_3,\vec{a}_1\otimes\vec{a}_3,\vec{a}_1\otimes\vec{a}_2)^T,\\
\vec{P}_3&=\vec{a}_1\otimes\vec{a}_2\otimes\vec{a}_3,
\end{align}
\label{eq:Pj}
\eml
map them to each other according to the transformation rules:
\beml\begin{align} U_\ur:&
\vec{P}_2\rightarrow \vec{P}_1^{\;\prime},\;\;
U_\llc:\vec{P}_1\rightarrow
\vec{P}_2^{\;\prime},\;\; U_\times: \vec{P}_2\rightarrow \vec{P}_2^{\;\prime},\\
W_\ur:& \vec{P}_3\rightarrow \vec{P}_2^{\;\prime},\;\;
W_\llc:\vec{P}_2\rightarrow \vec{P}_3^{\;\prime},\;\; W_\times:
\vec{P}_3\rightarrow \vec{P}_3^{\;\prime}.
\end{align}\label{mappingVW}\eml

Now, the action of the matrices $(U_{\lambda\mu})_\alpha$,
$(W_{\lambda\mu})_\alpha$,
($\lambda,\mu=1,2,3$) on the vectors $\vec{P}_j$ is fully determined by the action of the matrices $(\bar{v}_{\lambda\mu})_\alpha$ on the vectors $(\vec{p}_{\lambda\mu})_1$ and $(\vec{p}_{\lambda\mu})_2$. In other words, the explicit form of the vectors
$\vec{P}_j^{\;\prime}$ can obtained by appropriately constructing the three-atom vectors out of the vectors
$(\vec{p}_{\lambda\mu}^{\;\prime})_1$ and $(\vec{p}_{\lambda\mu}^{\;\prime})_2$, the latter vectors being defined in the right hand side of Eqs.~(\ref{eq:all_V}).

Vector $\vec{P}^{\prime}_1$ is created from the vector $(\vec{p}_{\lambda\mu}^{\;\prime})_1$  by appending the null vector $\vec{0}$ at the position corresponding to atom $\beta\neq\lambda\neq\mu$, such that the atomic indices of the resulting vector are arranged in decreasing order, as in the vector $\vec{P}_1$.
Vector $\vec{P}^{\prime}_2$ is created from the vector $(\vec{p}_{\lambda\mu}^{\;\prime})_2$ by appending
the 2 vectors $\vec{0}\otimes\vec{0}$, such that the atomic indices are arranged in the same order as in vector $\vec{P}_2$. Finally, the vector  $\vec{P}^{\prime}_3$ is created
form the vector $(\vec{p}_{\lambda\mu}^{\;\prime})_2$ by appending a tensor product with vector $\vec{a}_\beta$ ($\beta\neq\lambda\neq\mu$), such that the atomic indices are arranged in the same order as in the vector $\vec{P}_3$.

\section{Integral representations for the Green's functions}
\label{ap-Dx}
From the definitions of Eqs.~(\ref{eq:M_times}) and (\ref{eq:Greens}), it follows:
\be
G_\times(z)=G_{23}(z)\oplus G_{13}(z)\oplus G_{12}(z).
\label{newDefGx}
\e
Here, $G_{\lambda\mu}(z)$ is the Green's matrix for two independent atoms $\lambda$ and $\mu$:
\be
G_{\lambda\mu}(z)\equiv (z-M_{\lambda\mu})^{-1},
\e
where $M_{\lambda\mu}$ is the evolution matrix of two free atoms,
\be
M_{\lambda\mu}=M_\lambda\otimes\openone+\openone\otimes M_\mu.
\label{eq:Mij}
\e

As proven in Ref.~\cite{shatokhin10}, the Green's matrix
$G_{\lambda\mu}(z)$ can be represented as \be
G_{\lambda\mu}(z)=\int_{-\infty}^{\infty}\frac{d\omega'}{2\pi}G_\lambda\lt(\frac{z}{2}\pm
i\omega'\rt)\otimes G_\mu\lt(\frac{z}{2}\mp i\omega'\rt),
\label{eq:Gijz} \e where $G_\lambda(z)$ is defined in
Eq.~(\ref{eq:Greens}). Hence, $G_\times(z)$ (cf.
Eq.~(\ref{newDefGx})) is given by a direct sum of three integrals in
the form of Eq.~(\ref{eq:Gijz}).

Using this result, we will show that a similar integral representation can be obtained for the Green's matrix
\be
D_\times(z)=(z-N_\times)^{-1},\label{Dx}\e
governing evolution of three independent atoms.
Matrix $N_\times$, specified in Eq.~(\ref{eq:N_times}), can be written as
\be
N_\times=M_1\otimes \openone\otimes \openone+\openone\otimes M_{23}.
\e
Obviously, $N_\times$ has the same structure as the matrix $M_\times$, see Eq.~(\ref{eq:Mij}). Therefore, we can represent it, by analogy with Eq.~(\ref{eq:Gijz}), as
\be
D_\times(z)=\int_{-\infty}^{\infty}\frac{d\omega'}{2\pi} G_1\lt(\frac{z}{2}\pm i\omega'\rt)\otimes G_{23}\lt(\frac{z}{2}\mp i\omega'\rt).\label{eq:Dx-1}\e
Finally, using the integral representation for the Green's function $G_{23}(z')$:
\be
G_{23}(z')=\int_{-\infty}^{\infty}\frac{d\omega''}{2\pi}G_2\lt(\frac{z'}{2}\pm i\omega''\rt)
\otimes G_3\lt(\frac{z'}{2}\mp i\omega''\rt),
\e
where $z'\equiv z/2\mp i\omega'$, we obtain
\begin{align}
D_\times(z)&=\int_{-\infty}^{\infty}\int_{-\infty}^{\infty}\frac{d\omega'}{2\pi}\frac{d\omega''}{2\pi}G_1\lt(\frac{z}{2}\pm i\omega'\rt)\n\\
&\otimes G_2\lt(\frac{z'}{2}\pm i\omega''\rt)
\otimes G_3\lt(\frac{z'}{2}\mp i\omega''\rt),
\label{eq:Dx-2}
\end{align}
which yields the integral representation for the function $D_\times(z)$.

\section{Sum rules}
\label{sec:sum_rules}
The following sum rules constitute a method of calculating exactly sums of the frequency integrals appearing in the expressions for the vectors $\vec{x}^{\;(n)}$, $\vec{y}^{\;(n)}$, and $\vec{z}^{\;(n)}$ due to the integral representations
of the matrices $G_\times(z)$ and $D_\times(z)$ given by
Eqs.~(\ref{eq:Gijz}) and (\ref{eq:Dx-2}), respectively.

The idea of the method is based on the spectral decomposition of the single-atom
Green's matrices $G_\lambda(z)$, followed by calculation of the frequency integrals in the complex plane and, finally, a summation of the results. We will outline here the main steps in deriving the sum rules, and illustrate
them with two characteristic examples. Further useful, though restricted to the two-atom case, examples can be found in \cite{shatokhin10}.

We will first note that the Bloch matrix $M_\lambda$ governing the evolution
of atom $\lambda$ (see Eq.~(\ref{eq:BlochM})) allows for a spectral
decomposition \be M_\lambda=\sum_k r_kP_k^\lambda, \label{eq:decomp_M} \e
where \be P_k^\lambda=\frac{|u_k^\lambda\ra\la v_k^\lambda|}{\la v_k^\lambda|u_k^\lambda\ra},
\quad \sum_k P_k^\lambda=\openone, \e is the projector on the subspace
corresponding to the eigenvalue $r_k$, and $|u_k^\lambda\ra$
($|v_k^\lambda\ra$) are the right (left) eigenvectors of the matrix $M_\lambda$
satisfying $\la v_k^\lambda|u_{k'}^\lambda\ra=0$ for $k\neq k'$ \cite{horn}. The
eigenvalues $r_k$ of the matrix $M_\lambda$ are the roots of the
characteristic polynomial thereof: \be
f(z)=(z+2\gamma)\bigl((z+\gamma)^2+\delta^2\bigr)+(z+\gamma)|\Omega|^2.
\label{eq:charact_poly} \e While the explicit form of $r_k$
will be unimportant to us, we will rely on the fact that for
arbitrary $|\Omega|$ and $\delta$, the roots are non-degenerate, and
$\re[r_k]<0$.

To illustrate the sum rules, we will calculate the
zeroth-order correlation functions $\vec{y}^{\;(0)}$ and
$\vec{z}^{\;(0)}$. Although the results
are known without any calculations, cf. Eq.~(\ref{eq:init_cond}),
we will obtain them alternatively by using the sum rules.
These derivations highlight all steps
needed for tackling more complicated cases.

Taking $n=0$ in Eq.~(\ref{eq:y^n}), we obtain
\be
\vec{y}^{\;(0)}=G_\times L_+\vec{x}^{\;(0)}.\label{eq:y^0-1}
\e
Using Eqs. (\ref{eq:L_plus}), (\ref{newDefGx}),  and (\ref{eq:Gijz}), we can
bring Eq.~(\ref{eq:y^0-1}) to the form
\be \vec{y}^{\;(0)}=\lt({\cal G}_{23},{\cal G}_{13},{\cal
G}_{12}\rt)^T, \e where \begin{align} {\cal
G}_{\lambda\mu}&\equiv\int_{-\infty}^{\infty}\frac{d\omega'}{2\pi}
\lt(G_\lambda(i\omega')\vec{L}\otimes
G_\mu(-i\omega')G_\mu\vec{L}\rt.\n\\
&\lt.+G_\lambda(i\omega')G_\lambda\vec{L}\otimes
G_\mu(-i\omega')\vec{L}\rt), \label{eq:calG_ij} \end{align} and, for
definiteness, the upper sign in the integrals (\ref{eq:Gijz}) was
taken.

Next, inserting into (\ref{eq:calG_ij}) a spectral decomposition of
the single-atom Green's matrix \be
G_\lambda(z)=\sum_k\frac{P_k^\lambda}{z-r_k}, \label{ewq:decompG} \e
which follows directly from Eqs.~(\ref{eq:decomp_M}) and
(\ref{eq:Greens}), we arrive at
\begin{align}
{\cal G}_{\lambda\mu}&=\int_{-\infty}^{\infty}\frac{d\omega'}{2\pi}\n\\
&\times\sum_{k,l}\lt(\frac{P_k^\lambda}{i\omega'-r_k}\vec{L}\otimes \frac{P_l^\mu}{-i\omega'-r_l}
\frac{1}{-r_l}\vec{L}\rt.\n\\
&\lt.+\frac{P_k^\lambda}{i\omega'-r_k}\frac{1}{-r_k}\vec{L}\otimes
\frac{P_l^\mu}{-i\omega'-r_l}\vec{L}\rt). \label{eq:sum_1}
\end{align}
This integral can now easily be taken using the residues theorem. By
noting that, in the upper complex half-plane of $\omega'$, the
integrand of (\ref{eq:sum_1}) has simple poles at
$\omega'=-ir_k$, we obtain the result
\beq
{\cal G}_{\lambda\mu}&=&\sum_{k,l}\lt(\frac{P_k^\lambda}{1}\vec{L}\otimes \frac{P_l^\mu}{-r_k-r_l}
\frac{1}{-r_l}\vec{L}+\frac{P_k^\lambda}{-r_k}\vec{L}\otimes \frac{P_l^\mu}{-r_k-r_l}\rt)
\n\\
&=&\sum_{k,l}\frac{P_k^\lambda}{-r_k}\vec{L}\otimes
\frac{P_l^\mu}{-r_l}\vec{L} =G_\lambda\vec{L}\otimes G_\mu\vec{L},
\eq
whence, in agreement with Eq.~(\ref{eq:init_y^0}), it follows
\be \vec{y}^{\;(0)}=(G_2\vec{L}\otimes
G_3\vec{L},G_1\vec{L}\otimes G_3\vec{L}, G_1\vec{L}\otimes
G_2\vec{L})^T. \label{eq:y^0} \e

We will proceed with a calculation of $\vec{z}^{\;(0)}$ for which Eq.~(\ref{eq:z^n}) yields \be \vec{z}^{\;(0)}=D_\times
L_\times \vec{y}^{\;(0)}. \label{eq:z0} \e Inserting into
(\ref{eq:z0}) the integral representation (\ref{eq:Dx-2}) of
$D_\times\equiv D_\times(0)$, the definition of $L_\times$
(Eq.~(\ref{eq:nullblock})), as well as the result (\ref{eq:y^0}), we
get
\begin{widetext}
\beq
\vec{z}^{\;(0)}&=&\int_{-\infty}^{\infty}\int_{-\infty}^{\infty}
\frac{d\omega'}{2\pi}\frac{d\omega''}{2\pi}
\lt(G_1(i\omega')\vec{L}\otimes G_2(i\omega''_-)G_2\vec{L}
\otimes G_3(-i\omega''_+)G_3\vec{L}
+G_1(i\omega')G_1\vec{L}\otimes G_2(i\omega''_-)\vec{L}\otimes
G_3(-i\omega''_+)G_3\vec{L}\rt.\n\\
&&\lt.+G_1(i\omega')G_1\vec{L}
\otimes G_2(i\omega''_-)G_2\vec{L}\otimes G_3(-i\omega''_+)\vec{L}\rt)\equiv \int_{-\infty}^{\infty}\int_{-\infty}^{\infty}
\frac{d\omega'}{2\pi}\frac{d\omega''}{2\pi}{\cal G}_{123}(\omega',\omega''),
\label{eq:sumrule1}
\eq
with $\omega''_{\pm}=\omega''\pm \omega'/2$.
By virtue of the spectral decomposition (\ref{ewq:decompG}) the vector function
${\cal G}_{123}(\omega',\omega'')$ yields the expression
\begin{align}
{\cal
G}_{123}(\omega',\omega'')&=\sum_{k,l,m}\lt(\frac{P_k^1}{i\omega'-r_k}\vec{L}
\otimes
\frac{1}{i\omega''-i\omega'/2-r_l}\frac{P_l^2}{-r_l}\vec{L}
\otimes \frac{1}{-i\omega''-i\omega'/2-r_m}\frac{P_m^3}{-\lambda_m}\vec{L}\rt.\n\\
&\lt.+\frac{1}{i\omega'-r_k}\frac{P_k^1}{-r_k}\vec{L}
\otimes \frac{P_l^2}{i\omega''-i\omega'/2-r_l}\vec{L}
\otimes \frac{1}{-i\omega''-i\omega'/2-r_m}\frac{P_m^3}{-r_m}\vec{L}\rt.\n\\
&\lt.+\frac{1}{i\omega'-r_k}\frac{P_k^1}{-r_k}\vec{L}
\otimes
\frac{1}{i\omega''-i\omega'/2-r_l}\frac{P_l^2}{-r_l}\vec{L}
\otimes
\frac{P_m^3}{-i\omega''-i\omega'/2-r_m}\vec{L}\rt).
\label{eq:first_step_Z0}
\end{align}
Taking integral over $\omega''$ in the complex plane, we observe
that each of the lines of (\ref{eq:first_step_Z0}) has a simple pole
$\omega''=-ir_l+\omega'/2$ in the upper half-plane (and, also,
$\omega''=ir_m+\omega'/2$ -- in the lower half-plane), which
gives
\begin{align}
\int_{-\infty}^{\infty}\frac{d\omega''}{2\pi}{\cal G}_{123}
(\omega',\omega'')&=\sum_{k,l,m}\lt(\frac{P_k^1}{i\omega'-r_k}\vec{L}
\otimes \frac{P_l^2}{-r_l}\vec{L}
\otimes \frac{1}{-i\omega'-r_l-r_m}\frac{P_m^3}{-r_m}\vec{L}\rt.\n\\
&\lt.+\frac{1}{i\omega'-r_k}\frac{P_k^1}{-r_k}\vec{L}
\otimes \frac{P_l^2}{1}\vec{L}
\otimes \frac{1}{-i\omega'-r_l-r_m}\frac{P_m^3}{-r_m}\vec{L}\rt.\n\\
&\lt.+\frac{1}{i\omega'-r_k}\frac{P_k^1}{-r_k}\vec{L}
\otimes \frac{P_l^2}{-r_l}\vec{L} \otimes
\frac{P_m^3}{-i\omega'-r_l-r_m}\vec{L}\rt).
\label{eq:first_step_Z1}
\end{align}
Finally, the integral over $\omega'$ yields
\begin{align}
\vec{z}^{\;(0)}&=\sum_{k,l,m}\lt(\frac{P_k^1}{1}\vec{L} \otimes
\frac{P_l^2}{-r_l}\vec{L}
\otimes \frac{1}{-r_k-r_l-r_m}\frac{P_m^3}{-r_m}\vec{L}
+\frac{P_k^1}{-r_k}\vec{L} \otimes \frac{P_l^2}{1}\vec{L}
\otimes \frac{1}{-r_k-r_l-r_m}\frac{P_m^3}{-r_m}\vec{L}\rt.\n\\
&\lt.+\frac{P_k^1}{-r_k}\vec{L} \otimes
\frac{P_l^2}{-r_l}\vec{L}
\otimes \frac{P_m^3}{-r_k-r_l-r_m}\vec{L}\rt)
=\sum_{k,l,m}\frac{P_k^1}{-r_k}\vec{L}\otimes\frac{P_l^2}{-r_l}\vec{L}\otimes
\frac{P_m^3}{-r_m}\vec{L}
=G_1\vec{L}\otimes G_2\vec{L}\otimes G_3\vec{L},
\end{align}
which confirms Eq.~(\ref{eq:init_z0}).
\end{widetext}

It should be mentioned that the sum rules are valid also in a more
general case when the detunings and/or Rabi frequencies depend upon the atomic index $\lambda$. Indeed, the fact that the eigenvalues
$r_k^\lambda$ in these cases depend on $\lambda$ changes nothing in the
preceding derivations. That means that the results obtained in
this paper can be generalized to include, e.g., residual thermal
motion of atoms via individual Doppler shifts and, hence, detunings
$\delta_\lambda$, or wave propagation in the effective medium, in which
case the laser field strength (Rabi frequency) would depend on the
atomic position inside the medium.

\section{Forbidden diagrams}
\label{app:forbidden} \subsection{Combinations of diagrams from
Fig.~\ref{fig:expcross1}} \label{fa}
\begin{figure}
\includegraphics[width=2.5cm]{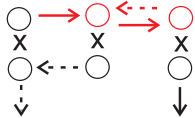}
\caption{(Color online) Diagram (c1)(f1)(d2) contains an amplitude
that comes in and cycles between the circles (highlighted with red)
without the outgoing arrow.}\label{fig:closed_loop}
\end{figure}
The following combinations are
forbidden: (c1)(f1)(d2), (c1)(f4)(d2), (c2)(f1)(d2), (c2)(f2)(d1),
(c2)(f2)(d2), (c2)(f2)(d3), (c2)(f3)(d2), (c2)(f4)(d1),
(c2)(f4)(d2), (c2)(f4)(d3), (c2)(f5)(d2), (c3)(f1)(d2),
(c3)(f4)(d2). All these diagrams as well as the ones listed below
contain closed loops (see an example in Fig.~\ref{fig:closed_loop}).

\subsection{Combinations of diagrams from Fig.~\ref{fig:expcross2}} \label{fb}
(a1)(g4)(d2), (a1)(g5)(d2), (a1)(g6)(d2), (a1)(g8)(d2),
(a2)(g4)(d2), (a2)(g5)(d2), (a2)(g6)(d2), (a2)(g8)(d2).


\begin{thebibliography}{99}
\bibitem{ftn1}
We will deal with the inelastic scattering processes
whereupon the outgoing wave number may differ from the incident one.
However, the associated relative changes are negligibly small and
are ignored throughout this work.
\end{thebibliography}
\end{document}